\documentclass[a4,11pt]{article}
\usepackage{amsfonts,latexsym,eucal,color,graphicx,epsfig,amssymb,amsmath,eucal,bm}

\newcommand{\be}{\begin{eqnarray}}
\newcommand{\ee}{\end{eqnarray}}
\newcommand{\beq}{\begin{eqnarray}}
\newcommand{\eeq}{\end{eqnarray}}
\newcommand{\pd}{\partial}

\newcommand{\nn}{\nonumber}

\makeatletter

\@addtoreset{equation}{section}
\makeatletter

\begin{document}

\begin{titlepage}

\begin{center}
{\Large {\bf Non-linear perturbation of black branes at large $D$}}\\

\vspace{3mm}

{\bf Umpei Miyamoto}\\
\vspace{1mm}
{\small {\it RECCS, Akita Prefectural University, Akita 015-0055, Japan}}\\
\vspace{1mm}
{\small {\tt umpei@akita-pu.ac.jp}}\\

\vspace{1mm}

\abstract{
The Einstein equations describing the black-brane dynamics both in Minkowski and AdS background were recently recast in the form of coupled diffusion equations in the large-$D$(imension) limit. Using such results in the literature, we formulate a higher-order perturbation theory of black branes in time domain and present the general form of solutions for arbitrary initial conditions. For illustrative purposes, the solutions up to the first or second order are explicitly written down for several kind of initial conditions, such as a Gaussian wave packet, shock wave, and rather general superposed sinusoidal waves. These could be  the first examples describing the non-trivial evolution of black-brane horizons in time domain. In particular, we learn some interesting aspects of black-brane dynamics such as the Gregory-Laflamme (GL) instability and non-equilibrium steady state (NESS). The formalism presented here would be applicable to the analysis of various black branes and their holographically dual field theories.
}
\end{center}

\tableofcontents

\end{titlepage}

\section{Introduction}

It is important to understand the dynamics of higher-dimensional black objects, since it tells us much about the nature of higher-dimensional gravitational theories and their holographically dual quantum field theories. The strong non-linearity of gravity, however, usually prevents us from understanding the dynamical properties of black objects beyond the linear-perturbation regime without highly sophisticated skills of numerical computation.

The Gregory-Laflamme (GL) instability~\cite{Gregory:1993vy}, which is a universal instability of higher-dimensional black objects, is a good example to see the above situation. Though the GL instability in the non-linear regime is quite interesting, its analysis needs sophisticated skills of numerical relativity~\cite{Lehner:2010pn}. While there exists a semi-analytic higher-order perturbation method~\cite{Gubser:2001ac}, it seems applicable only to static problems.

Recently, Emparan, Suzuki, and Tanabe showed that the Einstein equations describing the horizon dynamics of black branes in both Minkowski and Anti-de Sitter (AdS) background are recast in the form of coupled non-linear diffusion-type equations when the number of spatial dimensions is large~\cite{Emparan:2015gva}. This result provides us with a unique approach to the non-linear dynamics of black objects in higher dimensions. The authors indeed showed that the unstable black strings converge to non-uniform black strings (NUBSs), which had been predicted to happen above a critical dimension~\cite{Sorkin:2004qq}, by solving the diffusion equations numerically with a few lines of {\sl Mathematica} code. It is added that the blackfold approach~\cite{Camps:2010br} is also thought to serve as a powerful approach to analyze the evolution of GL instability. 

Once the simple diffusion equations were obtained~\cite{Emparan:2015gva}, it is natural to ask if the non-linear properties of black-brane dynamics can be understood analytically. In this paper, we develop a systematic non-linear perturbation theory of asymptotically flat and AdS black branes, allowing the perturbations to be dynamical. Using the Fourier and Laplace transformation to solve the partial differential equations (PDEs), the perturbation equations are solved order by order for given arbitrary  initial conditions up to the integration associated with the inverse transformation.

While the formulation is so general that it would be applicable to various problems, we pick up several examples as the initial conditions, which are a Gaussian wave packet, a step-function like shock configuration, and quite general discretely superposed sinusoidal waves. For these examples, the integration associated with the inverse transformation is completed up to the first or second order, and the properties of solutions are examined. Through these examples, one will see the validity of formalism itself and some unknown, or yet-to-be-confirmed, non-linear properties of black-brane dynamics. In particular, in the case of asymptotically flat black branes, an interesting non-liner property of GL instability resulting from the mode-mode coupling is unveiled at the second order. In the case of shock propagation on asymptotically AdS black branes, the analytic description of non-equilibrium steady state (NESS), which was recently discussed in the Riemann problem of relativistic fluid mechanics and field theories~\cite{Herzog:2016hob}, is presented. 

This paper is organized as follows. In Sec.~\ref{sec:flat}, the asymptotically flat black branes are investigated. In Sec.~\ref{sec:formalism}, we present the perturbation equations for asymptotically flat black branes and their general form of solutions. In Sec.~\ref{sec:gauss}, we apply the general result to the Gaussian wave packet. In Sec.~\ref{sec:sin}, we consider the discretely superposed sinusoidal waves. In Sec.~\ref{sec:ads}, we consider the non-linear perturbation of asymptotically AdS black branes. Here, the formulation and applications are presented in parallel with Sec.~\ref{sec:flat}, but a new example of initial condition, the step-function like shock, is investigated in Sec.~\ref{sec:shock}. Section~\ref{sec:conc} is devoted to conclusion. Throughout this paper, we follow the notations in Ref.~\cite{Emparan:2015gva}.

\section{Asymptotically flat black branes}
\label{sec:flat}

\subsection{Perturbation equations and general form of solutions}
\label{sec:formalism}

In the large-$D$(imension) approach, the horizon dynamics of vacuum black branes without a cosmological constant are described by two functions, $m(t,z)$ and $p(t,z)$, where $t$ is time and $z$ is the spatial coordinate along which the horizon extends~\cite{Emparan:2015gva}. $m$ and $p$ represent the mass and momentum distributions along the horizon, respectively. $m \to +0$ corresponds to the pinching off of the horizon. The equations of motion for these quantities take form of coupled non-linear diffusion equations,
\begin{gather}
	(\pd_t - \pd_z^2) m  + \pd_z p
	=
	0,
\label{eom1}
\\
	(\pd_t - \pd_z^2) p - \pd_z m
	=
	- \pd_z ( \frac{p^2}{m} ),
\label{eom2}
\\
	t>0, 
\;\;\;
	-\infty < z < \infty.
\label{domain}
\end{gather}

A uniform black-brane solution corresponds to $m(t,z) \equiv  1$ and $p(t,z) \equiv 0$. Since we are interested in the dynamical deformation of such a uniform solution, we introduce one-parameter families of $m(t,z)$ and $p(t,z)$, and expand them around the uniform black-brane solution,
\begin{gather}
	m(t,z;\epsilon) = 1+ \sum_{\ell=1}^\infty m_\ell (t,z) \epsilon^\ell,
\label{expansion1}
\\
	p(t,z;\epsilon) =  \sum_{\ell=1}^\infty p_\ell (t,z) \epsilon^\ell,
\label{expansion2}
\end{gather}
where $\epsilon$ is a constant parameterizing the families. Substituting these expansions into Eqs.~\eqref{eom1} and \eqref{eom2}, we obtain the equations of motion at $ O(\epsilon^\ell) \; (\ell \in {\mathbb N})$,
\begin{gather}
	\dot{m}_\ell - m_\ell'' + p_\ell' 
	=
	0,
\label{peom1}
\\
	\dot{p}_\ell - p_\ell'' -m_\ell' 
	=
	\psi_\ell,
\label{peom2}
\end{gather}
where the dot and prime denote the derivatives with respect to $t$ and $z$, respectively. The right-hand side of Eq.~\eqref{peom2}, $\psi_\ell (t,z)$, which we call a source term, is a polynomial of the lower-order perturbations and their first spatial derivatives,
\begin{gather}
	\psi_1 \equiv 0,
\label{source1}
\\
	\psi_\ell
	=
	\psi_\ell ( m_1, p_1, m_1', p_1', \cdots , m_{\ell-1}, p_{\ell-1}, m_{\ell-1}', p_{\ell-1}'  ),
\;\;\;
	\ell \geq 2.
\label{source>1}
\end{gather}
For example, the source terms for $\ell=2$ and $\ell =3$ are given by
\begin{gather}
	\psi_2
	=
	-2p_1 p_1',
\label{source2}
\\
	\psi_3 
	=
	2 m_1 p_1 p_1' + m_1' p_1^2  -2 p_1 p_2' - 2p_1' p_2.
\label{source3}
\end{gather}

In the rest of this section, we are looking for the general form of solutions to the perturbation equations~\eqref{peom1} and \eqref{peom2}, combining the Fourier and Laplace transformations (see, {\it e.g.}, \cite{Duffy}). A similar technique is found to be used in Ref.~\cite{CFM,Miyamoto:2008uf} to analyze the higher-order perturbation of surface-diffusion equation, which is a single non-linear PDE.

Before starting to solve Eqs.~\eqref{peom1} and \eqref{peom2}, let us introduce the notations associated with the Fourier and Laplace transformations. For a given function, say $f(t,z)$, we shall denote its Fourier transformation with respect to $z$ by $\bar{f}(t,k)$, and its Laplace transformation with respect to $t$ by the corresponding capital letter $F(s,z)$. Namely,
\begin{align}
	\bar{f}(t,k) &:= {\cal F}[ f(t,z) ] = \int_{-\infty}^\infty f(t,z) e^{-ik z} dz,
\;\;\;
	i:=\sqrt{-1},
\\
	F(s,z)  &:= {\cal L}[f(t,z)] = \int_0^\infty f(t,z) e^{-st} dt.
\end{align}
Then, a capital letter with a bar denotes a Fourier-Laplace transformation as 
\be
\bar{F}(s,k) :=  ( {\cal L} \circ {\cal F} ) [ f(t,z) ]. 
\ee
In addition, we define two kind of convolutions, 
\begin{gather}
	f(t,z)  \ast g(t,z) := \int_0^t f(t-\tau,z) g(\tau,z) d\tau,
\\
	f(t,z) \star g   (t,z) := \int_{-\infty}^{\infty} f(t,z-\xi) g(\tau,\xi) d\xi.
\end{gather}

With the notations introduced above, the Fourier-Laplace transformed version of Eqs.~\eqref{peom1} and \eqref{peom2} are written as coupled algebraic equations in a matrix form
\begin{gather}
	{\bm A}
	\left(
	\begin{array}{c}
		\bar{M}_\ell(s,k) \\
		\bar{P}_\ell(s,k) \\
	\end{array}
	\right)
	=
	\left(
	\begin{array}{c}
		\bar{m}_\ell (0,k) \\
		\bar{p}_\ell (0,k) + \bar{\Psi}_\ell(s,k) \\
	\end{array}
	\right),
\label{peom3}
\\
	{\bm A}
	:=
	\left(
	\begin{array}{cc}
		s+k^2  & ik \\
		- ik  & s+k^2 \\
	\end{array}
	\right),
\label{A}
\end{gather}
where we have used ${\cal F}[ \pd_z^n f(t,z) ] = (ik)^n \bar{f}(t,k) \; (n \in {\mathbb N})$ and $ {\cal L}[ \pd_t f(t,z) ] = sF(s,z) -f(0,z)$.

\begin{figure}[tb]
\begin{center}
\begin{minipage}[c]{0.9\textwidth}
\linespread{0.85}
\begin{center}
\includegraphics[height=4cm]{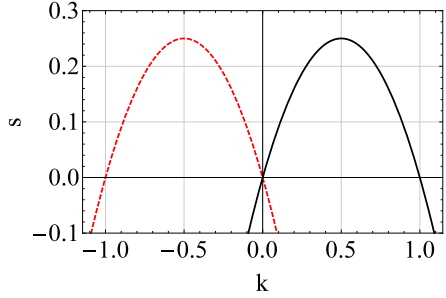}
\caption{{\small {\sf Dispersion relations $s=s_+(k)$ (black solid) and $s=s_-(k)$ (red dashed), defined by Eq.~\eqref{dispersion}. 
}}}
\label{fig:disp}
\end{center}
\end{minipage}
\end{center}
\end{figure}

The solution to Eqs.~\eqref{peom1} and \eqref{peom2} are obtained after multiplying Eq.~\eqref{peom3} by ${\bm A}^{-1}$ from left and inversely transforming it,
\be
	\left(
	\begin{array}{c}
		m_\ell(t,z) \\
		p_\ell(t,z) \\
	\end{array}
	\right)
	=
	({\cal F}^{-1} \circ {\cal L}^{-1})
	\left[
	{\bm A}^{-1}
	\left(
	\begin{array}{c}
		\bar{m}_\ell (0,k) \\
		\bar{p}_\ell (0,k) + \bar{\Psi}_\ell(s,k) \\
	\end{array}
	\right)
	\right] .
\label{sol}
\ee
By simple algebra, the inverse matrix ${\bm A}^{-1}$ is found to be decomposed into two parts,
\begin{gather}
	{\bm A}^{-1}
	=
	\sum_{\sigma = +,-} \frac{ 1 }{s-s_\sigma (k)} {\bm B}_\sigma,
\label{Ainverse}
\\
	{\bm B}_\sigma
	:=
	\frac12
	\left(
	\begin{array}{cc}
		1 &  - \sigma i \\
		\sigma i & 1\\
	\end{array}
	\right),
\label{B}
\\
	s_\sigma (k) := k(\sigma 1 - k ).
\label{dispersion}
\end{gather}
See Fig.~\ref{fig:disp} for plot of $s=s_\pm(k)$, which corresponds to the dispersion relation of waves.
After this decomposition, one can perform the inverse Laplace transformation ${\cal L}^{-1}$ in Eq.~\eqref{sol} to obtain
\begin{align}
	\left(
	\begin{array}{c}
		m_\ell(t,z) \\
		p_\ell(t,z) \\
	\end{array}
	\right)
	&=
	\sum_{\sigma=+,-}
	{\bm B}_\sigma
	\left(
	\begin{array}{c}
		{\cal F}^{-1} [ e^{ s_\sigma (k) t } \bar{m}_\ell (0,k) ]\\
		{\cal F}^{-1} [ e^{ s_\sigma (k) t } \bar{p}_\ell (0,k) + e^{ s_\sigma (k) t } \ast \bar{ \psi }_\ell (t,k) ]\\
	\end{array}
	\right)
\label{sol2}
\\
	&=
	\sum_{\sigma=+,-}
	{\bm B}_\sigma
	\left(
	\begin{array}{c}
		{\cal F}^{-1}[ e^{s_\sigma (k) t}  ] \star m_\ell (0,z) \\
		{\cal F}^{-1}[ e^{s_\sigma (k) t}  ] \star p_\ell (0,z) + {\cal F}^{-1}[ e^{s_\sigma (k) t}  ] \star \ast \psi_\ell (t,z)  \\ 
	\end{array}
	\right).
\label{sol3}
\end{align}
Here, we have used ${\cal L}^{-1}[ \frac{1}{s-a} ] = e^{a t}$, ${\cal L}^{-1}[ F(s,z) G(s,z) ] = f(t,z) \ast g(t,z)$, and ${\cal F}^{-1}[ \bar{f}(t,k) \bar{g}(t,k) ] = f(t,z) \star g(t,z)$.

Equations \eqref{sol2} and \eqref{sol3} are exactly what we wanted, namely, the general form of solutions to the perturbation equations~\eqref{peom1} and \eqref{peom2} for given arbitrary initial conditions, $ m_\ell (0,z) $ and $ p_\ell (0,z) \; (\ell \in {\mathbb N})$. It depends on the problem which expression, \eqref{sol2} or \eqref{sol3}, is easer to compute. For all examples considered in this paper, Eq.~\eqref{sol2} seems easer to compute. Using Eq.~\eqref{dispersion}, one can easily obtain
\be
	{\cal F}^{-1}[ e^{s_\sigma (k) t}  ] = \frac{ 1 }{ \sqrt{4\pi t} } \exp[ - \frac{ (\sigma t+iz)^2}{4t} ] ,
\label{fourier_exp}
\ee
which is useful when one uses expression \eqref{sol3}.

In principle, one can obtain the arbitrary-order solutions, $m_\ell (t,z)$ and $p_\ell(t,z)$, by computing the right-hand side of Eq.~\eqref{sol2} or \eqref{sol3} order by order. However, as the source term $\psi_\ell (t,z) $ becomes complicated as $\ell$ increases, it is fare to say that to obtain the solutions analytically until arbitrary order is impossible in general. In addition, when the initial condition is a complicated function, even the first-order solution can be impossible to obtain analytically. Namely, the inverse Fourier transformation in Eq.~\eqref{sol2} cannot be performed analytically in such a case. In the rest of this section, we shall consider two examples of initial conditions, for which the first few-order solutions are analytically obtainable.

\subsection{Gaussian wave packet}
\label{sec:gauss}

As the first example, we adopt the Gaussian wave packet as the initial perturbation given to the asymptotically flat black brane. For this perturbation, the inverse Fourier transformation at the first order in Eqs.~\eqref{sol2} and \eqref{sol3} can be computed analytically. Since the unperturbed black brane is unstable, the perturbation of course grows unboundedly and nothing unexpected happens in this sense. However, one can see how to use the general results obtained in the previous section and the validity of the method. In particular, comparing the perturbative solution with a full numerical solution, the perturbative solution turns out to be effective even for finite-amplitude dynamics. In other words, the convergence of $\epsilon$-expansion \eqref{expansion1} and \eqref{expansion2} is rapid enough at least for this example.

Let us assume that the situation where the black brane is given $O(\epsilon)$-perturbation taking form of a Gaussian wave packet,
\begin{gather}
	m_1(0,z) = \frac{ \beta }{ \sqrt{2\pi} b } \exp [ - \frac{ (z-z_0)^2 }{ 2b^2 } ],
\label{ic_gauss1}
\\
	p_1 (0,z) = m_1'(0,z),
\label{ic_gauss2}
\end{gather}
where $\beta,\; b \; (>0)$, and $z_0$ are real constants. Obviously, $b$ and $z_0$ parameterize how much the wave packet spatially extends and the central position of the wave packet, respectively. $\beta$ is just a normalization constant to give $ \int_{-\infty}^\infty m_1(0,z) dz = \beta $.  The initial perturbation of $p_1$ is given by the spatial derivative of $m_1$ for simplicity, though their initial conditions can be independent in nature.

The Fourier transformation of the above initial conditions are
\begin{gather}
	\bar{m}_1(0,k)
	=
	\beta \exp [ -\frac{ b^2 k^2 }{2}  - ik z_0 ],
\\
	\bar{p}_1(0,k) = ik \bar{m}_1(0,k).
\end{gather}

\subsubsection{First-order solutions}

Since we have no source term at $O(\epsilon)$, $\psi_1(t,z) \equiv 0$, we see from Eq.~\eqref{sol2} that what to compute is the inverse Fourier transformation of initial spectra, $\bar{m}_1(0,k)$ and $\bar{p}_1(0,k)$, multiplied by $e^{s_\sigma(k) t}$. One can compute such quantities as
\begin{align}
	{\cal F}^{-1}[ e^{s_\sigma (k) t}  \bar{m}_1(0,k)]
	&=
	\frac{ \beta }{ \sqrt{ 2\pi ( b^2+2t )   } } 
	\exp
		[
		\frac{ t^2-( z-z_0 )^2 }{ 2( b^2+2t ) }
		+
		i\sigma \frac{  t ( z-z_0 ) }{ b^2+2t }
	],
\\
	{\cal F}^{-1}[ e^{s_\sigma (k) t}  \bar{p}_1(0,k)]
	&=
	- \frac{ \beta [ (z-z_0) - i \sigma t   ] }{ \sqrt{ 2\pi ( b^2+2t )^3 }    }  
	\exp
		[
		\frac{ t^2-( z-z_0 )^2 }{ 2( b^2+2t ) }
		+
		i\sigma \frac{  t ( z-z_0 ) }{ b^2+2t }
		].
\end{align}
Substituting these results into Eq.~\eqref{sol2}, we obtain the first-order solutions
\begin{align}
	\left(
	\begin{array}{c}
		m_1 (t,z) \\
		p_1 (t,z) \\
	\end{array}
	\right)
	=
	\beta \sqrt{ \frac{ (b^2+3t)^2+(z-z_0)^2 }{  2\pi (b^2+2t)^3 } } 
	\exp [ \frac{ t^2-( z-z_0 )^2 }{ 2( b^2+2t ) } ] 
	\left(
	\begin{array}{c}
		\displaystyle \cos  [\frac{ t(z-z_0) }{ b^2+2t } + \Theta ]  \\
		\displaystyle  - \sin [ \frac{ t(z-z_0) }{ b^2+2t } + \Theta ] \\
	\end{array}
	\right),
\label{mp1_gauss}
\end{align}
where
\begin{align}
	\cos \Theta
	:=
	\frac{ b^2+3t }{ \sqrt{ (b^2+3t)^2+(z-z_0)^2 } },
\;\;\;
	\sin \Theta
	:=
	\frac{ z-z_0 }{ \sqrt{ (b^2+3t)^2+(z-z_0)^2 } }.
\end{align}

We believe that this is the first example describing the non-trivial evolution of GL instability analytically in time domain, which is realized by virtue of the large-$D$ method and perturbation theory developed in this paper.

Like the diffusion phenomenon of a Gaussian wave packet according to an ordinary diffusion equation, solutions \eqref{mp1_gauss} have temporal decay factor, which behaves as $  \frac{1}{\sqrt{ b^2+2t }} $, and spatial decay factor $\exp[ - \frac{ (z-z_0)^2 }{ 2(b^2+2t) } ]$. The sinusoidal parts represent spatial oscillation with time-dependent wavelength $ \frac{2\pi (b^2+2t)}{t} $, which interestingly asymptotes to a universal value $4\pi$ as $t \to +\infty$. 

What crucially different from the ordinary diffusion is that the solutions temporally grow exponentially due to factor $ \exp[ \frac{ t^2 }{ 2(b^2+2t) } ] $. 
It is stressed that this exponential growing happens eventually irrespective of $b$, which characterizes the extension of the initial wave packet. Substituting $z=z_0$ into Eq.~\eqref{mp1_gauss}, we can see the time dependence of the peak height,
\be
	m_1(t,z_0)
	=
	\beta \sqrt{ \frac{ (b^2+3t)^2 }{  2\pi (b^2+2t)^3 } }  \exp[ \frac{t^2}{2(b^2+2t)} ].
\ee
For $ b >\sqrt{3}$, this is monotonically increasing in time. For $ 0<b<\sqrt{3} $, although it is initially decreasing, it turns increasing at $t = \frac13 ( 3-2b^2 + \sqrt{ b^4-3b^2+9 } )$ to diverge eventually. 

Thus, while the GL instability is generally said to be a long-wavelength instability, the initial perturbation taking form of a Gaussian wave packet necessarily grows exponentially however the `scale' of perturbation $b$ is small. The reason is that the Fourier spectrum of any Gaussian wave packet necessarily contains the GL mode $k \in (-1,1) \setminus \{ 0 \}$ (see Sec.~\ref{sin_1st}). This is quite reasonable but might be a somewhat interesting point.

\begin{figure}[tb]
\begin{center}
\begin{minipage}[c]{0.9\textwidth}
\linespread{0.85}
\begin{center}
		\setlength{\tabcolsep}{ 10 pt }
		\begin{tabular}{ cc }
			(a) & (b) \\
			\includegraphics[height=4.5cm]{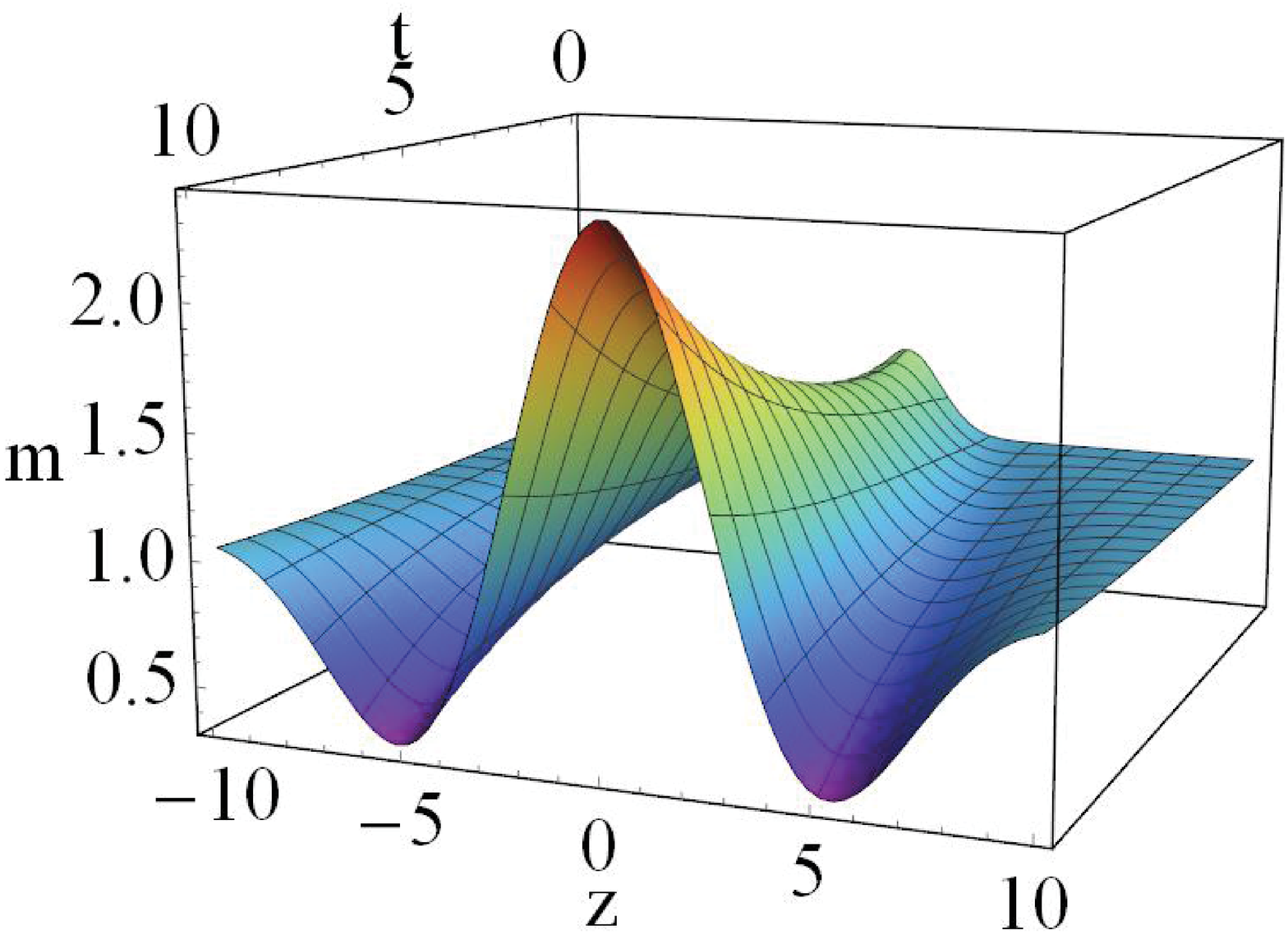} &
			\includegraphics[height=4.5cm]{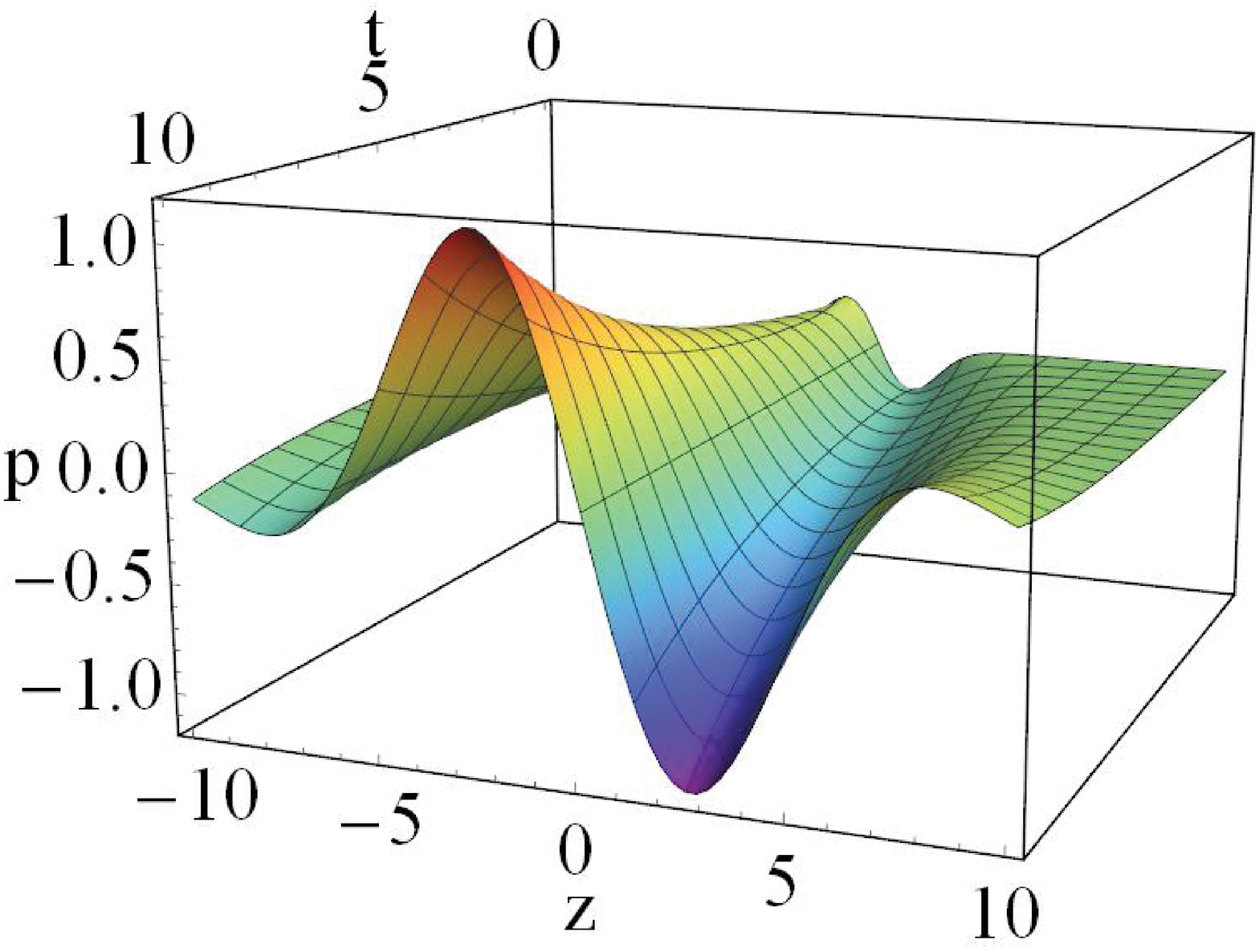} \\
			(c) & (d) \\
			\includegraphics[height=4cm]{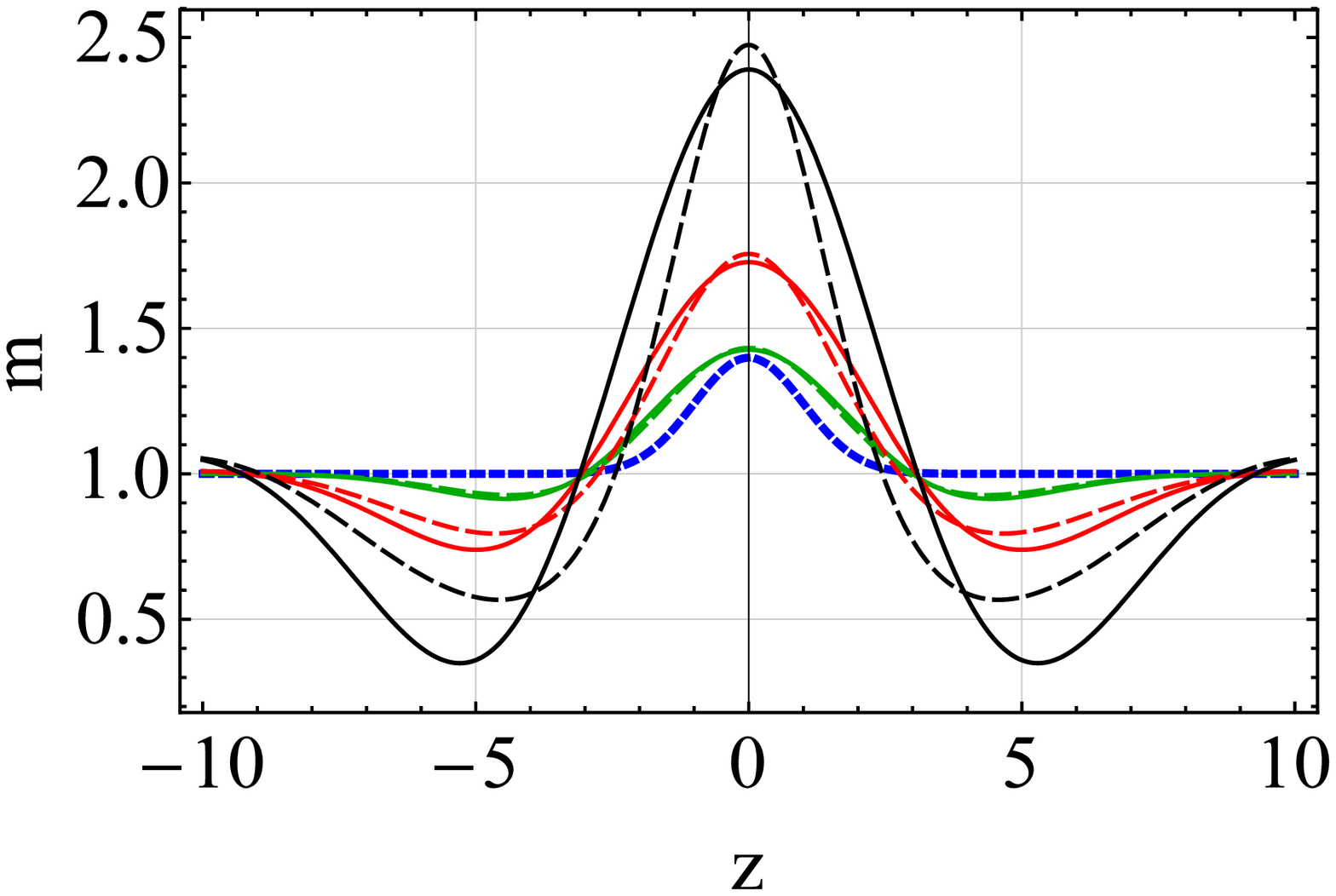} &
			\includegraphics[height=4cm]{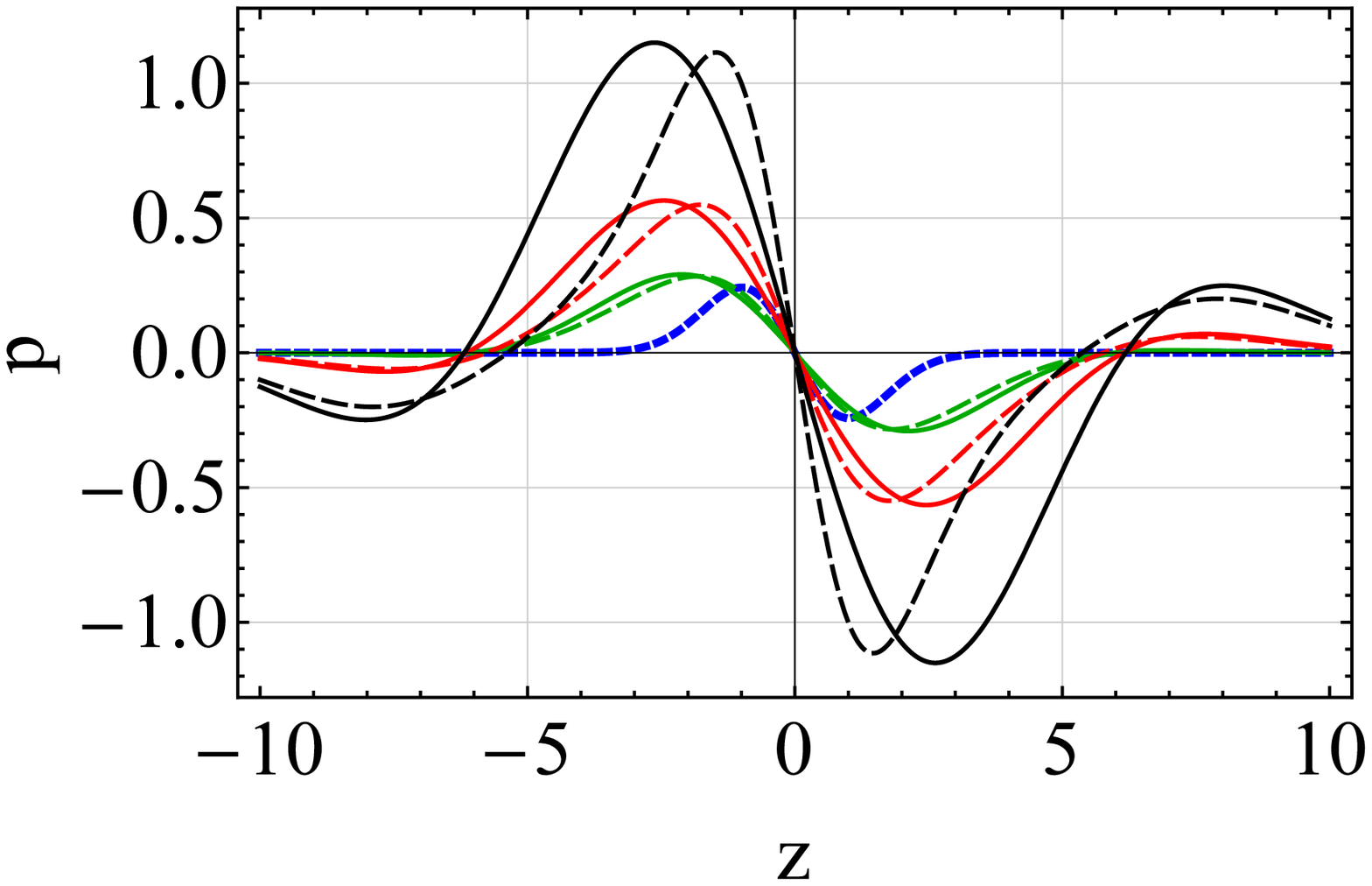} \\
		\end{tabular}
\caption{{\small {\sf Three-dimensional plots of (a) $1+m_1(t,z)$ and (b) $p_1(t,z)$, given by Eqs.~\eqref{mp1_gauss} with $\beta=b=1, \; z_0=0$. The comparison between first-order solution (c) $1+m_1(t,z)$ (resp.\ (d) $p_1(t,z)$) and non-perturbative solution $m(t,z)$ (resp.\ $p(t,z)$) obtained by numerically solving Eq.~\eqref{eom1} and \eqref{eom2}. The blue-dashed curve represents initial configuration $m(0,z)$ (resp.\ $p(0,z)$). The green, red, and black solid curves represent the first-order solutions at $t=3.3$, $6.7$, and $10$, respectively. The (green, red, and black) dashed curves represent the full numerical solutions at the corresponding time.}}}
\label{fig:gauss}
\end{center}
\end{minipage}
\end{center}
\end{figure}

Three-dimensional plots of $1+m_1(t,z)$ and $p_1(t,z)$ are presented in Figs.~\ref{fig:gauss}(a) and \ref{fig:gauss}(b), respectively. One can observe the growth and oscillation described above. In addition, snapshots of $1+m_1(t,z)$ and $p_1(t,z)$ at selected moments, compared with numerical solutions, are presented in Figs.~\ref{fig:gauss}(c) and \ref{fig:gauss}(d), respectively. Compared with the full numerical solutions, which are obtained by directly solving original equations \eqref{eom1} and \eqref{eom2} with the same initial conditions, {\it i.e.}, $m(0,z)=1+m_1(0,z)$ and $p(0,z)=p_1(0,z)$, one can observe that the first-order solutions almost completely capture the qualitative features of the full solution during the time domain considered. Note that the deviation from the full solution, however, becomes large as the time proceeds, which results in the divergence of $m$ and $p$.

\subsubsection{Notes on second-order perturbation}
\label{sec:note_gauss}

The comparison between the first-order solution and full solution above tells us that $O(\epsilon^2)$ perturbations are negligible during the amplitudes of $m$ and $p$ are $O(1)$ in the current example despite the ordinary expectation that the perturbation becomes invalid for such a large amplitude. We will see in Sec.~\ref{sec:ads} that $O(\epsilon)$ approximation is more accurate for the Gaussian perturbation to the asymptotically AdS brane than the present case.

For the aim to see the non-linear effects at the second order, it is natural to assume that the initial perturbation at the second order vanishes, $m_2(0,z)=p_2(0,z)=0$. The reason is that $m_2(t,z)$ and $p_2(t,z)$  are composed of two independent parts as seen in Eq.~\eqref{sol2}: one is the contribution from initial perturbation $m_2(0,z)$ and $p_2(0,z)$, and the other is that from the source term $\psi_2(t,z)$.  The former clearly has the same time dependence as the first term from Eq.~\eqref{sol2}. Namely, if we prepare the Gaussian wave packet as the initial condition of second-order perturbation, the second-order perturbation evolves in the exactly same way as the $O(\epsilon)$ perturbation described above. Only the latter, the contribution from the source term, can have a different time dependence from the first-order perturbation. This will be seen explicitly in Sec.~\ref{sec:sin}.

From the reason described above, we should assume that the initial perturbations vanish at $O(\epsilon^2)$, $m_2(0,z)=p_2(0,z)=0$. Then, we see from Eq.~\eqref{sol2} that what to compute at $O(\epsilon^2)$ is only the inverse Fourier transformation of the convolution between $e^{s_\sigma(k)t}$ and the spectrum of source term $\bar{\psi}_2 (t,k)$. Unfortunately, however, such a convolution in the present example involves the Gauss error function, not written in terms of elementary functions. Thus, it seems difficult to obtain the second-order solutions analytically, and therefore we stop the analysis on this example here.

\subsection{Superposed sinusoidal waves}
\label{sec:sin}

As the second example, we consider the situation where the black brane is initially given an $O(\epsilon)$-perturbation being a superposition of an arbitrary number of sinusoidal waves. This example is simple but interesting enough to see what happens in the non-linear regimes.

We set the following initial conditions
\begin{gather}
	m_1 (0,z) =\sum_{n=1}^N a_n \cos (k_n z + \varphi_n),
\label{ic1}
\\
	p_1 (0,z) = m_1'(0,z),
\label{ic2}
\\
	 m_\ell (0,z) = p_\ell (0,z) = 0,
\;\;\;
	\forall \ell \geq 2,
\label{ic3}
\end{gather}
where $a_n,$ $k_n$, and $\varphi_n \; (n=1,2,\cdots, N)$ are real constants. It is noted that the right-hand side of Eq.~\eqref{ic1} is not written as the general form of Fourier series expansion of a periodic function. However, choosing appropriate wave number $k_n$ and phase $\varphi_n$, and taking summation over $n$ from $0$ to infinity, rather than from 1 to $N$, Eq.~\eqref{ic1} can cover the Fourier series expansion of arbitrary piecewise continuous periodic function. The assumption that the second and higher order perturbations vanish initially are adopted from the same reason as the previous example in Sec.~\ref{sec:gauss}.

The Fourier transformations of the above initial configurations are
\begin{gather}
	\bar{m}_1 (0,k)
	=
	\pi \sum_{n=1}^N a_n
	[ e^{ i\varphi_n } \delta (k-k_n) + e^{- i \varphi_n } \delta (k+k_n) ],
\\
	\bar{p}_1(0,k) = ik \bar{m}_1(0,k),
\\
	\bar{m}_\ell (0,k) = \bar{p}_\ell (0,k) = 0,
\;\;\;
	\forall \ell \geq 2.
\end{gather}
In the rest of this section, we shall compute the right-hand side of Eq.~\eqref{sol2} order by order for these initial conditions.

\subsubsection{First-order solutions}
\label{sin_1st}

Since we have no source term at $O(\epsilon)$, $ \psi_1 \equiv 0 $, we see from Eq.~\eqref{sol2} that what to compute is only the inverse Fourier transformation of the initial spectra, $\bar{m}_1(0,k)$ and $\bar{q}_1 (0,k)$, multiplied by $ e^{s_\sigma (k) t} $. These are easily computed to give
\begin{align}
	{\cal F}^{-1}[ e^{ s_\sigma (k) t } \bar{m}_1 (0,k) ]
	=&
	\frac12 \sum_{n=1}^N a_n
	[
		e^{ s_\sigma (k_n) t } e^{ i( k_n z + \varphi_n ) }
		+
		e^{ s_{- \sigma} (k_n) t } e^{ - i( k_n z + \varphi_n ) }
	],
\label{Fem}
\\
	{\cal F}^{-1}[ e^{ s_\sigma (k) t } \bar{p}_1 (0,k) ]
	=&
	\frac{i}{2} \sum_{n=1}^N k_n a_n
	[
		e^{ s_\sigma (k_n) t } e^{ i( k_n z + \varphi_n ) }
		-
		e^{ s_{- \sigma} (k_n) t } e^{ - i( k_n z + \varphi_n ) }
	],
\label{Fep}
\end{align}
where we have used $s_\sigma (-k) = s_{-\sigma} (k)$. Substituting these results into Eq.~\eqref{sol2}, we obtain the first-order solutions,
\begin{align}
	m_1(t,z)
	=&
	\frac12 \sum_{n=1}^N a_n
	[
		(1 + k_n) e^{ s_+ (k_n) t }  + (1 - k_n) e^{ s_- (k_n) t }
	] \cos ( k_n z + \varphi_n ),
\label{m1}
\\
	p_1(t,z)
	=&
	- \frac12 \sum_{n=1}^N a_n
	[
		(1 + k_n) e^{ s_+ (k_n) t }  - (1 - k_n) e^{ s_- (k_n) t }
	] \sin ( k_n z + \varphi_n ).
\label{p1}
\end{align}

Equations~\eqref{m1} and \eqref{p1} represent $O(\epsilon)$ approximate time evolution of the initial perturbation, which takes the form of superposed sinusoidal waves~\eqref{ic1} and \eqref{ic2}. Since the initial conditions \eqref{ic1} and \eqref{ic2} are quite general, so are the solutions \eqref{m1} and \eqref{p1}.

Since we are considering linear equations of motion, there is no mode-mode coupling appearing in non-linear regime, and therefore Eqs.~\eqref{m1} and \eqref{p1} have simple interpretation. The factor of $\cos(k_n z+\varphi_n)$ in Eq.~\eqref{m1} represents the time-dependent amplitude of the initially given mode $\cos(k_n z+\varphi_n)$. Each mode evolves independently according to its growth or damping rate determined by $e^{s_+(k_n)t}$ and $e^{s_-(k_n)t}$. From the concrete form of $s_\pm(k)$ in Eq.~\eqref{dispersion}, one can see that if $k_n \in (-1,0)$ (resp.\ $k_n \in (0,1)$), such a mode grows exponentially due to $e^{s_-(k_n)t}$ (resp.\ $e^{s_+ (k_n)t}$), which represents the GL instability.

\subsubsection{Second-order solutions}

Since we assume that the initial perturbations vanish at $O(\epsilon^2)$, $ m_2(0,z)=p_2(0,z)=0 $, we see from Eq.~\eqref{sol2} that what to compute is the inverse Fourier transformation of the convolution between $ e^{ s_\sigma (k) t } $ and the Fourier spectrum of source term $\bar{\psi}_2(t,k)$. Using Eqs.~\eqref{source2} and \eqref{p1}, such a quantity is computed and written down in a simple form as
\begin{align}
	{\cal F}^{-1}[ e^{ s_\sigma (k) t } \ast & \bar{ \psi }_2 (t,k) ]
	=
	\frac{i}{8}
	\sum_{n=1}^N \sum_{n'=1}^N a_n a_{n'} k_{n'}
\nn
\\
&
\times
	\Big(
		C_{nn'}^{(\sigma)(+)} e^{i[ (k_n+k_{n'})z+(\varphi_n + \varphi_{n'}) ] } 
		+
		C_{nn'}^{(\sigma)(-)} e^{i[ (k_n-k_{n'})z+(\varphi_n - \varphi_{n'}) ]} 
\nn
\\
&
		-
		C_{nn'}^{(-\sigma)(-)} e^{-i[ (k_n-k_{n'})z+(\varphi_n - \varphi_{n'}) ]} 
		-
		C_{nn'}^{(-\sigma)(+)} e^{-i[ (k_n+k_{n'})z+(\varphi_n + \varphi_{n'}) ]} 
	\Big)
\label{convo}
\end{align}
by defining a function of time,
\begin{align}
	C_{nn' }^{(\sigma)(\sigma') } 
	:=&
	\frac{ ( 1+k_n )( 1+k_{n'} ) }{ s_+(k_n) + s_+(k_{n'}) - s_\sigma ( k_n + \sigma' k_{n'} )}
	( e^{ [ s_+(k_n) + s_+(k_{n'}) ] t } - e^{ s_\sigma (k_n +\sigma' k_{n'}) t } )
\nn
\\
	-&
	\frac{ ( 1+k_n )( 1-k_{n'} ) }{ s_+(k_n) + s_-(k_{n'}) - s_\sigma ( k_n +\sigma' k_{n'} )}
	( e^{ [ s_+(k_n) + s_-(k_{n'}) ] t } - e^{ s_\sigma (k_n +\sigma' k_{n'}) t } )
\nn
\\
	-&
	\frac{ ( 1-k_n )( 1+k_{n'} ) }{ s_-(k_n) + s_+(k_{n'}) - s_\sigma( k_n +\sigma' k_{n'} )}
	( e^{ [ s_-(k_n) + s_+(k_{n'}) ] t } - e^{ s_\sigma (k_n +\sigma' k_{n'}) t } )
\nn
\\
	+&
	\frac{ ( 1-k_n )( 1-k_{n'} ) }{ s_-(k_n) + s_-(k_{n'}) - s_\sigma ( k_n +\sigma' k_{n'} )}
	( e^{ [ s_-(k_n) + s_-(k_{n'}) ] t } - e^{ s_\sigma (k_n +\sigma' k_{n'}) t } ).
\label{c}
\end{align}
Substituting the above result \eqref{convo} into Eq.~\eqref{sol2}, we obtain the second-order solutions,
\begin{align}
m_2(t,z)
	=
	\frac18 \sum_{ n =1 }^n \sum_{n'=1}^N a_n a_{n'} k_{n'}
	\Big(
		[ & C_{nn'}^{(+)(+)} - C_{nn'}^{(-)(+)} ] 
		\cos [ (k_n+k_{n'})z + (\varphi_n + \varphi_{n'}) ]
\nn
\\
	+&
	 [ C_{nn'}^{(+)(-)} - C_{nn'}^{(-)(-)} ]  \cos [ (k_n-k_{n'})z + (\varphi_n - \varphi_{n'}) ]
	\Big), 
\label{m2_pre}
\\
p_2(t,z)
	=
	- \frac18 \sum_{ n =1 }^n \sum_{n'=1}^N a_n a_{n'} k_{n'}
	\Big(
		[ & C_{nn'}^{(+)(+)} + C_{nn'}^{(-)(+)} ] 
		\sin [ (k_n+k_{n'})z + (\varphi_n + \varphi_{n'}) ]
\nn
\\
	+&
	 [ C_{nn'}^{(+)(-)} + C_{nn'}^{(-)(-)} ]  \sin [ (k_n-k_{n'})z + (\varphi_n - \varphi_{n'}) ]
	\Big) .
\label{p2_pre}
\end{align}

Note that $m(t,z) = 1+m_1(t,z)+m_2(t,z)$ and $p(t,z) = p_1(t,z)+p_2(t,z)$ with Eqs.~\eqref{m1}, \eqref{p1}, \eqref{m2_pre}, and \eqref{p2_pre}, represent $O(\epsilon^2)$ approximate time evolution of the initial perturbation, which takes the form of superposed sinusoidal waves~\eqref{ic1}, \eqref{ic2}, and \eqref{ic3}. Since the initial conditions \eqref{ic1} and \eqref{ic2} are quite general, so are these approximate solutions.

Since the initial perturbations are assumed to vanish at $O(\epsilon^2)$, $m_2(0,z)=p_2(0,z)=0$, the above $O(\epsilon^2)$ solutions contain only the contribution from the source term $\psi_2=-2p_1p_1'$. If one prepares for non-vanishing initial conditions at the second order, its contribution is simply added to the above solution, but such a contribution will exhibit no interesting behavior since it has the time dependence similar to that of $O(\epsilon)$ solution.

In general, the multiple summation of any quantity with two indices $\sum_{n,n'} T_{nn'}$ can be decomposed as $\sum_{n,n'} T_{nn'} = \sum_{n} T_{nn} + \sum_{n<n'} ( T_{nn'} +T_{n'n} )$. Here, $  \sum_{n<n'} $ represents the summation over all $n$ and $n'$ satisfying $1 \leq n<n' \leq N$.  Using this decomposition, one can decompose the multiple summation in Eqs.~\eqref{m2_pre} and \eqref{p2_pre} as
\begin{align}
&	m_2(t,z)
\nn
	=
\\
	& \frac18 \sum_{n=1}^N a_n^2 k_n
	\Big(
	[ C_{nn}^{(+)(+)} - C_{nn}^{ (-)(+) } ] \cos [ 2(k_n z+\varphi_n) ]
	+
	[ C_{nn}^{(+)(-)} - C_{nn}^{ (-)(-) } ]
	\Big)
\nn
\\
	+&
	\frac18 \sum_{ n < n' } a_n a_{n'}
	\Big(
		k_{n'}[ C_{nn'}^{(+)(+)} - C_{nn'}^{(-)(+)} ] + k_{n}[ C_{n'n}^{(+)(+)} - C_{n'n}^{(-)(+)} ]
	\Big) \cos [ (k_n+k_{n'})z + (\varphi_n + \varphi_{n'}) ]
\nn
\\
	+&
	\frac18 \sum_{ n < n' } a_n a_{n'}
	\Big(
		k_{n'}[ C_{nn'}^{(+)(-)} - C_{nn'}^{(-)(-)} ] + k_{n}[ C_{n'n}^{(+)(-)} - C_{n'n}^{(-)(-)} ]
	\Big) \cos [ (k_n-k_{n'})z + (\varphi_n - \varphi_{n'}) ],
\label{m2}
\\
&
	p_2(t,z)
	=
\nn
\\
&
	-\frac18 \sum_{n=1}^N a_n^2 k_n [ C_{nn}^{(+)(+)} + C_{nn}^{(-)(+)} ] \sin [ 2(k_n z+\varphi_n) ]
\nn
\\
&
	-\frac18 \sum_{n<n'} a_n a_{n'}
	\Big(
		k_{n'} [ C_{nn'}^{(+)(+)} + C_{nn'}^{(-)(+)} ] + k_{n} [ C_{n'n}^{(+)(+)} + C_{n'n}^{(-)(+)} ] 
	\Big)
	\sin [ (k_n+k_{n'}) z +( \varphi_n+\varphi_{n'} ) ]
\nn
\\
&
	-\frac18 \sum_{n<n'} a_n a_{n'}
	\Big(
		k_{n'} [ C_{nn'}^{(+)(-)} + C_{nn'}^{(-)(-)} ] - k_{n} [ C_{n'n}^{(+)(-)} + C_{n'n}^{(-)(-)} ] 
	\Big)
	\sin [ (k_n - k_{n'}) z +( \varphi_n - \varphi_{n'} ) ].
\label{p2}
\end{align}
The first term of Eqs.~\eqref{m2} and \eqref{p2} represents the self-interference of each mode $k_n$ ($ n = 1,2,\cdots, N $). On the other hand, the second and third terms represent the interference between $k_n$ and $k_{n'}$ ($n < n'$).

The non-linear source term involves the mode-mode coupling, which is absent at the linear order. This coupling excites the terms of $\cos[(k_n \pm k_{n'})z]$ and $\sin[(k_n \pm k_{n'})z]$ in Eqs.~\eqref{m2_pre} and \eqref{p2_pre}.  For example, let us see the structure of $m_2(t,z)$. From Eqs.~\eqref{c} and \eqref{m2_pre}, one can see that both $\cos[(k_n + k_{n'})z]$ and $\cos[(k_n - k_{n'})z]$ terms involve the following three kind of time dependence,
\begin{gather}
	 e^{ [ s_+(k_n) + s_+(k_{n'}) ] t },
\;\;\;
	e^{ [ s_+(k_n) + s_-(k_{n'}) ] t} ,
\;\;\;
	e^{ [ s_-(k_n) + s_-(k_{n'}) ] t } .
\label{factor1}
\end{gather}
In addition, one can see that $\cos[(k_n + k_{n'})z]$ and $\cos[(k_n - k_{n'})z]$ terms involve
\begin{gather}
	 e^{ s_+ (k_n + k_{n'}) t },
\;\;\;
	e^{ s_- (k_n + k_{n'}) t }
\;\;\;
\mbox{and}
\;\;\;
	e^{ s_+ (k_n - k_{n'}) t },
\;\;\;
	e^{ s_- (k_n - k_{n'}) t },
\label{factor2}
\end{gather}
respectively. Thus, the second-order solutions exhibit a variety of dispersion given by the exponents of quantities \eqref{factor1} and \eqref{factor2}.

\begin{figure}[bt]
	\begin{center}
\begin{minipage}[c]{0.9\textwidth}
\linespread{0.85}
\begin{center}
		\setlength{\tabcolsep}{ 10 pt }
		\begin{tabular}{ cc }
				(a) & (b) \\
			\includegraphics[height=4.5cm]{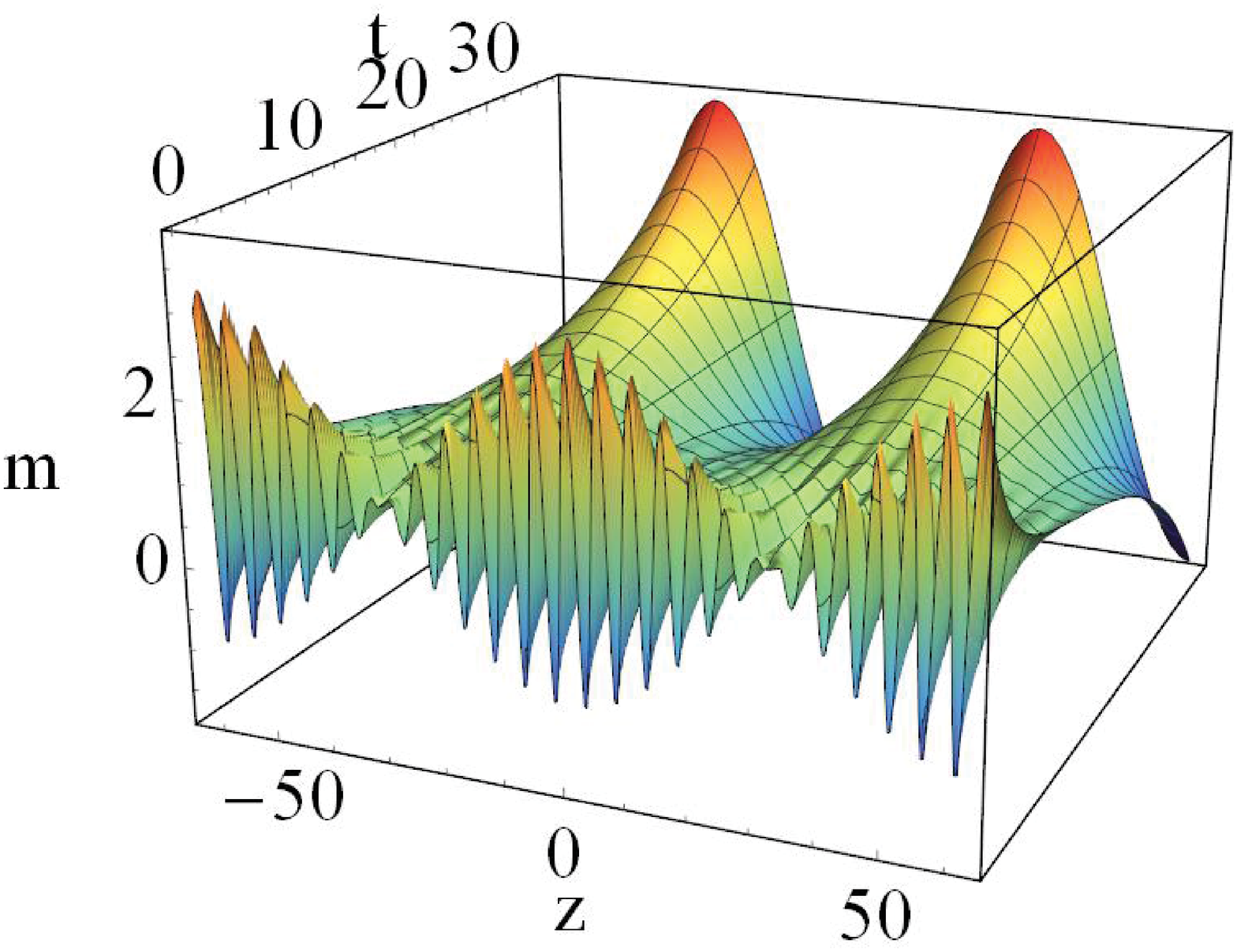} &
			\includegraphics[height=4.5cm]{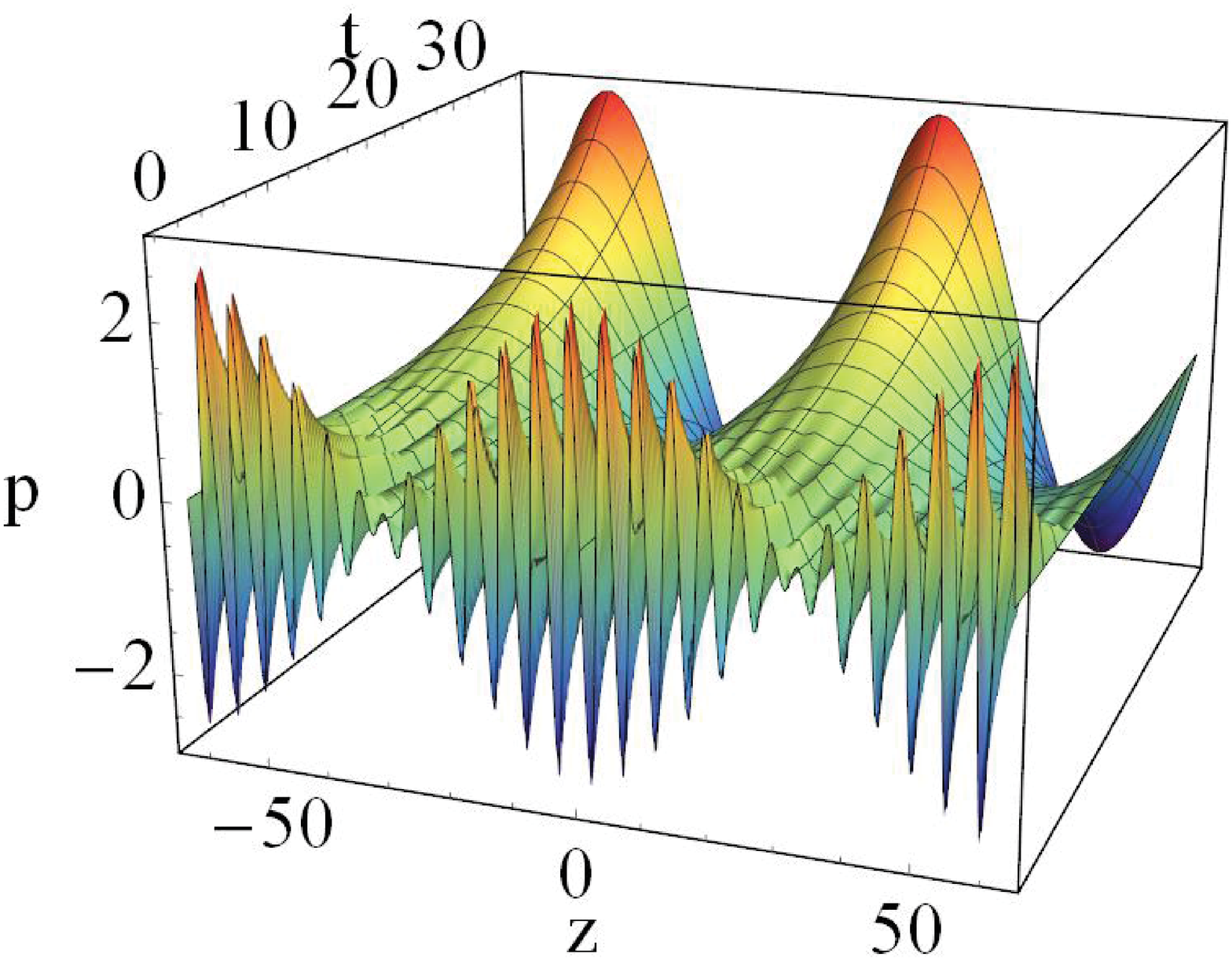} \\
				(c) & (d) \\
			\includegraphics[height=4cm]{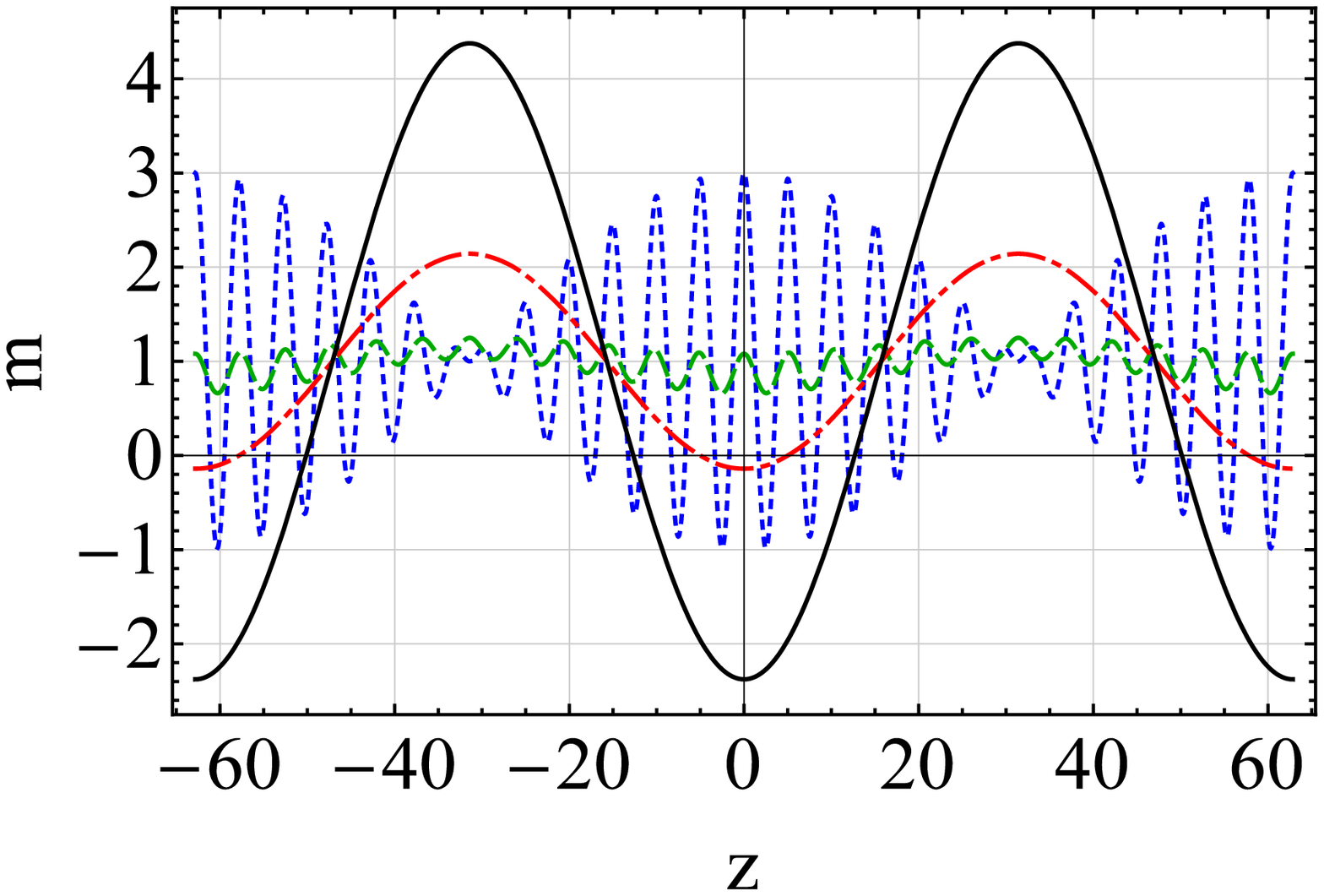} &
			\includegraphics[height=4cm]{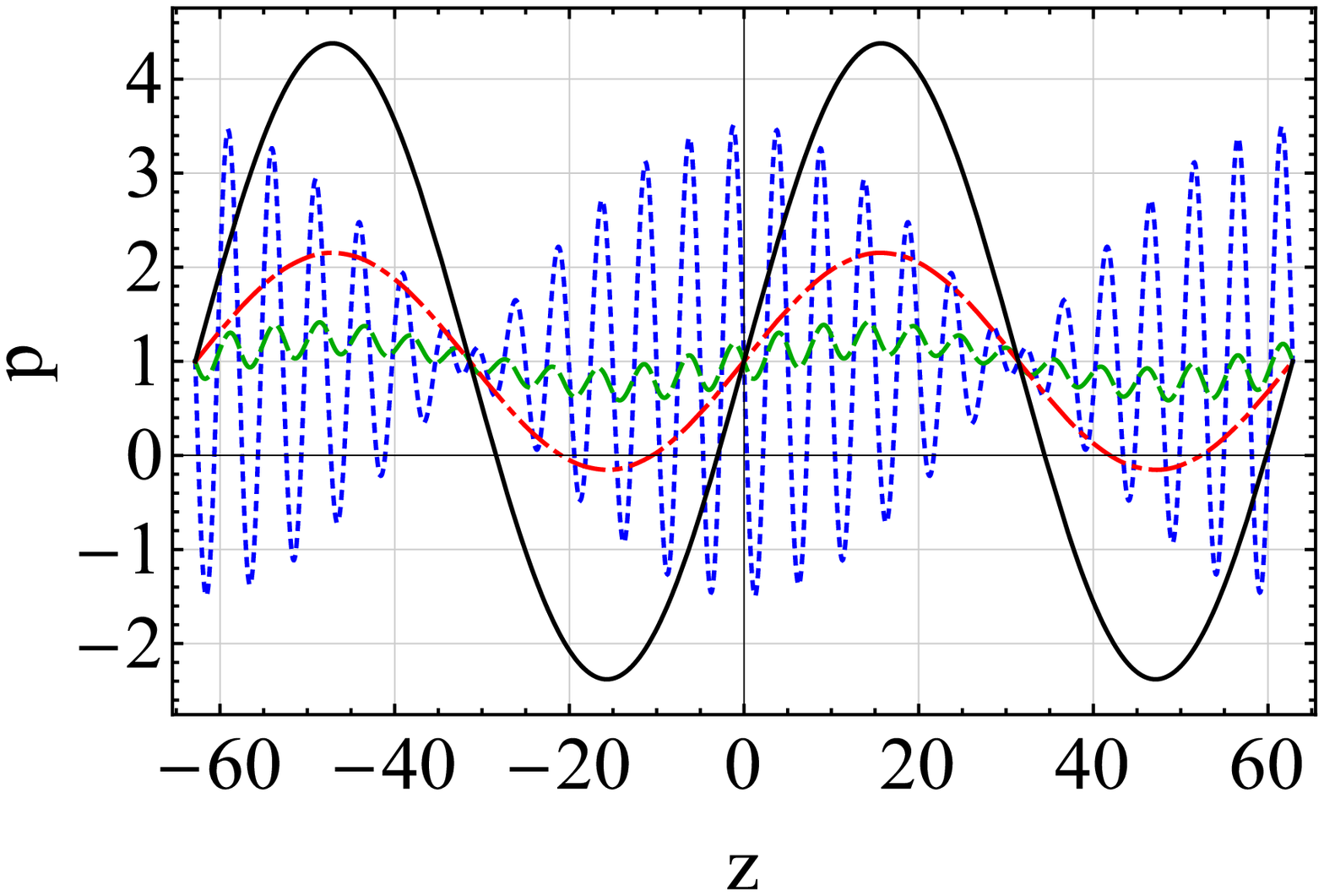} \\
		\end{tabular}
	\caption{{\small {\sf Three-dimensional plots of (a) $1+m_1(t,z)+m_2(t,z)$ and (b) $p_1(t,z)+p_2(t,z)$, given by Eqs.~\eqref{m1}, \eqref{p1}, \eqref{m2}, and \eqref{p2} with $N=2, \; a_1=a_2=1, \; k_1 = 1.3, \; k_2=1.2, \; \varphi_1=\varphi_2=0$. Snapshots of (c) $1+m_1(t,z)+m_2(t,z)$ and (d) $p_1(t,z)+p_2(t,z)$ at $t=0$ (blue dotted), $t=8$ (green dashed), $t=28$ (red dot-dashed), and $t=40$ (black solid).}}}
	\label{fig:sin}
\end{center}
\end{minipage}
	\end{center}
\end{figure}

\subsubsection{Notes on Gregory-Laflamme instability}

Let us consider the meaning to investigate the higher-order perturbations from the stability point of view. The asymptotically flat black brane we consider here is essentially unstable. Namely, as seen in Sec.~\ref{sin_1st}, if the initial perturbation contains any mode of which wave number $k_n \in (-1,1) \setminus \{ 0 \}$, such a mode grows unboundedly. However, we consider the black brane in the large-$D$ limit, namely, above the critical dimension~\cite{Sorkin:2004qq}. Thus, the GL instability initially grows but it gradually damps in non-linear regimes, and eventually the horizon converges to non-uniform configuration~\cite{Emparan:2015gva}.

It is pointed out that the second-order perturbation cannot stabilize the first-order instability because the first-order perturbation is treated as the fixed background, which appears as the source term, when we solve the second order. Nevertheless, one might expect if there appears the sign of the convergence to the non-uniform horizon at the second order. Although such a sign can appear at the second order, we cannot catch the sign in the result \eqref{m2} and \eqref{p2} unfortunately.

On the other hand, a black brane that is linearly stable can become unstable at the second order. If the initial perturbation does not contain any unstable mode, the initial perturbation will damp exponentially at linear level, as seen in the results of Sec.~\ref{sin_1st}. However, the second-order perturbation involves various time dependence as seen in Eqs.~\eqref{factor1} and \eqref{factor2}. In order to see directly this situation, let us focus on a simple case as follow. 

Suppose that the initial perturbation is the superposition of two modes $k_1$ and $k_2$ both of which are stable modes, $k_1> k_2 > 1$. In addition, assume that their difference is smaller than unity, $k_1 - k_2 \in (0,1)$. In this case, the term of $ C^{(+)(-)}_{21} \cos[ (k_1-k_2)z +(\varphi_1-\varphi_2)] $ in Eq.~\eqref{m2} includes terms having growing factor $ e^{s_+(k_1-k_2)t} $, as seen from Eq.~\eqref{c}. Thus, the perturbation does not grow at $O(\epsilon)$ but does at $O(\epsilon^2)$.

The above phenomenon is an interesting aspect of the GL instability, which was revealed for the first time by the present non-linear perturbation theory in time domain. It is intuitively understandable. The superposition of the two modes forms the beat. For simplicity, assume $ a_1=a_2 \; (\neq 0)$ in Eq.~\eqref{ic1}, then the superposed wave is written as
\be
	2 a_1 \cos [ \frac{ (k_1+k_2)z + (\varphi_1+\varphi_2) }{2} ]
		  \cos [ \frac{ (k_1-k_2)z + (\varphi_1-\varphi_2) }{2} ].
\ee
This exhibits the fast spatial oscillation with the large wave number $\frac{k_1+k_2}{2}$ which is enveloped by the slow oscillation with the small wave number $\frac{k_1-k_2}{2}$, which is called the beat phenomenon especially when the difference of the wave numbers is rather small $k_1-k_2 \ll k_1+k_2 $. This slow oscillation is nothing but the origin of the GL instability at the second order. In Fig.~\ref{fig:sin}, we present the three-dimensional plots of $m(t,z) = 1 + m_1(t,z) + m_2(t,z)$ and $ p_1(t,z) + p_2(t,z) $ and their snapshots at selected moments. One can observe that the beat formed by the superposition of two modes at $t=0$. As soon as the dynamics starts, such an initial wave rapidly damps as predicted by the first-order perturbation. As the time proceeds, however, the waves of which scale is the same order as that of the beat begin to grow and eventually diverge.

\section{Asymptotically AdS black branes}
\label{sec:ads}

In this section, we consider the non-linear perturbation of the asymptotically AdS black branes in the large-$D$ limit. The governing equations of motion are almost the same as those in the asymptotically flat case except for a sign of one term. Thus, the formulation of the perturbation theory proceeds completely in parallel with the asymptotically flat case. 

What different from the asymptotically flat black brane in the previous section is that the AdS black branes are stable: they are not suffered from the GL instability at least linearly. In addition, the gravitational phenomena in AdS background are interpreted as corresponding phenomena in the dual field theories via the AdS/CFT dictionary, and therefore have much more applications than the asymptotically flat case. In fact, we will apply the result of general perturbation theory to the problem of shock-wave propagation, which has been discussed in the context of AdS/CFT~\cite{Herzog:2016hob}, in Secs.~\ref{sec:shock} and \ref{sec:sin_ads} in addition to the Gaussian wave packet and general superposed sinusoidal waves.

\subsection{Perturbation equations and general form of solutions}
\label{sec:formalism_ads}

For the asymptotically AdS neutral black branes in the large-$D$ limit of general relativity, the mass and momentum distribution, $m(t,z)$ and $p(t,z)$, obey Eq.~\eqref{eom1} and the following equation~\cite{Emparan:2015gva,Herzog:2016hob} with the domain~\eqref{domain},
\be
	(\pd_t - \pd_z^2 ) p  + \pd_z m = -\pd_z ( \frac{p^2}{m} ).
\label{eom2_ads}
\ee
Substituting the expansion~\eqref{expansion1} and \eqref{expansion2} into Eqs.~\eqref{eom1} and \eqref{eom2_ads}, we obtain Eq.~\eqref{peom1} and
\be
	\dot{p}_\ell - p_\ell'' + m_\ell' = \psi_\ell,
\label{expansion2_ads}
\ee
where the source term $\psi_\ell$ is the same as those in the asymptotically Minkowski case, Eqs.~\eqref{source1}--\eqref{source3}.

Performing the Laplace and Fourier transformation in Eqs.~\eqref{peom1} and \eqref{expansion2_ads}, we obtain a couple of algebraic equations, which is written as Eq.~\eqref{peom3} with ${\bm A}$ replaced by the following matrix,
\be
	{\bm D}
	:=
	\left(
	\begin{array}{cc}
		s+k^2 & ik \\
		ik & s+k^2 \\
	\end{array}
	\right).
\label{D}
\ee
The inverse matrix ${\bm D}^{-1}$ is decomposed into two parts,
\begin{gather}
	{\bm D}^{-1}
	=
	\sum_{\sigma=+,-} \frac{1}{s-{\mathsf s}_\sigma (k)} {\bm E}_\sigma,
\label{Dinverse}
\\
	{\bm E}_\sigma
	:=
	\frac12
	\left(
		\begin{array}{cc}
			1 & -\sigma 1 \\
			-\sigma 1 & 1 \\
		\end{array}
	\right),
\label{E}
\\
	{\mathsf s}_\sigma (k) := k (\sigma i - k).
\label{dispersion_ads}
\end{gather}
After this decomposition, one can perform the inverse Laplace transformation ${\cal L}^{-1}$ in Eq.~\eqref{sol} with ${\bm A}$ replaced by ${\bm D}$. Then, the general form of solution is given by
\begin{align}
	\left(
	\begin{array}{c}
		m_\ell(t,z) \\
		p_\ell(t,z) \\
	\end{array}
	\right)
	&=
	\sum_{\sigma=+,-}
	{\bm E}_\sigma
	\left(
	\begin{array}{c}
		{\cal F}^{-1} [ e^{ {\mathsf s}_\sigma (k) t } \bar{m}_\ell (0,k) ]\\
		{\cal F}^{-1} [ e^{ {\mathsf s}_\sigma (k) t } \bar{p}_\ell (0,k) + e^{ {\mathsf s}_\sigma (k) t } \ast \bar{ \psi }_\ell (t,k) ]\\
	\end{array}
	\right)
\label{sol2_ads}
\\
	&=
	\sum_{\sigma=+,-}
	{\bm E}_\sigma
	\left(
	\begin{array}{c}
		{\cal F}^{-1}[ e^{{\mathsf s}_\sigma (k) t}  ] \star m_\ell (0,z) \\
		{\cal F}^{-1}[ e^{{\mathsf s}_\sigma (k) t}  ] \star p_\ell (0,z) + {\cal F}^{-1}[ e^{{\mathsf s}_\sigma (k) t}  ] \star \ast \psi_\ell (t,z)  \\ 
	\end{array}
	\right).
\label{sol3_ads}
\end{align}

While it depends on the chosen initial condition which expression between \eqref{sol2_ads} and \eqref{sol3_ads} is easier to compute, Eq.~\eqref{sol2_ads} is solely used in the rest of this paper. Using Eq.~\eqref{dispersion_ads}, the inverse Fourier transformation of $ e^{{\mathsf s}_\sigma (k) t } $ is easily computed as 
\be
	{\cal F}^{-1} [ e^{{\mathsf s}_\sigma (k) t} ] 
	=
	\frac{1}{ \sqrt{ 4\pi t } } \exp[ -\frac{( \sigma t+z )^2 }{ 4t } ].
\label{fourier_exp_ads}
\ee
This will be useful when one uses expression~\eqref{sol3_ads}.

\subsection{Gaussian wave packet}
\label{sec:gauss_ads}

As in Sec.~\ref{sec:gauss}, we investigate the Gaussian wave packet as the initial perturbation given to the asymptotically AdS black brane. Compared with the full-order numerical solution, the linear-order approximation turns out to be rather accurate approximation in this case.

\begin{figure}[tb]
	\begin{center}
\begin{minipage}[c]{0.9\textwidth}
\linespread{0.85}
\begin{center}
		\setlength{\tabcolsep}{ 10 pt }
		\begin{tabular}{ cc }
			(a) & (b) \\
			\includegraphics[height=4.5cm]{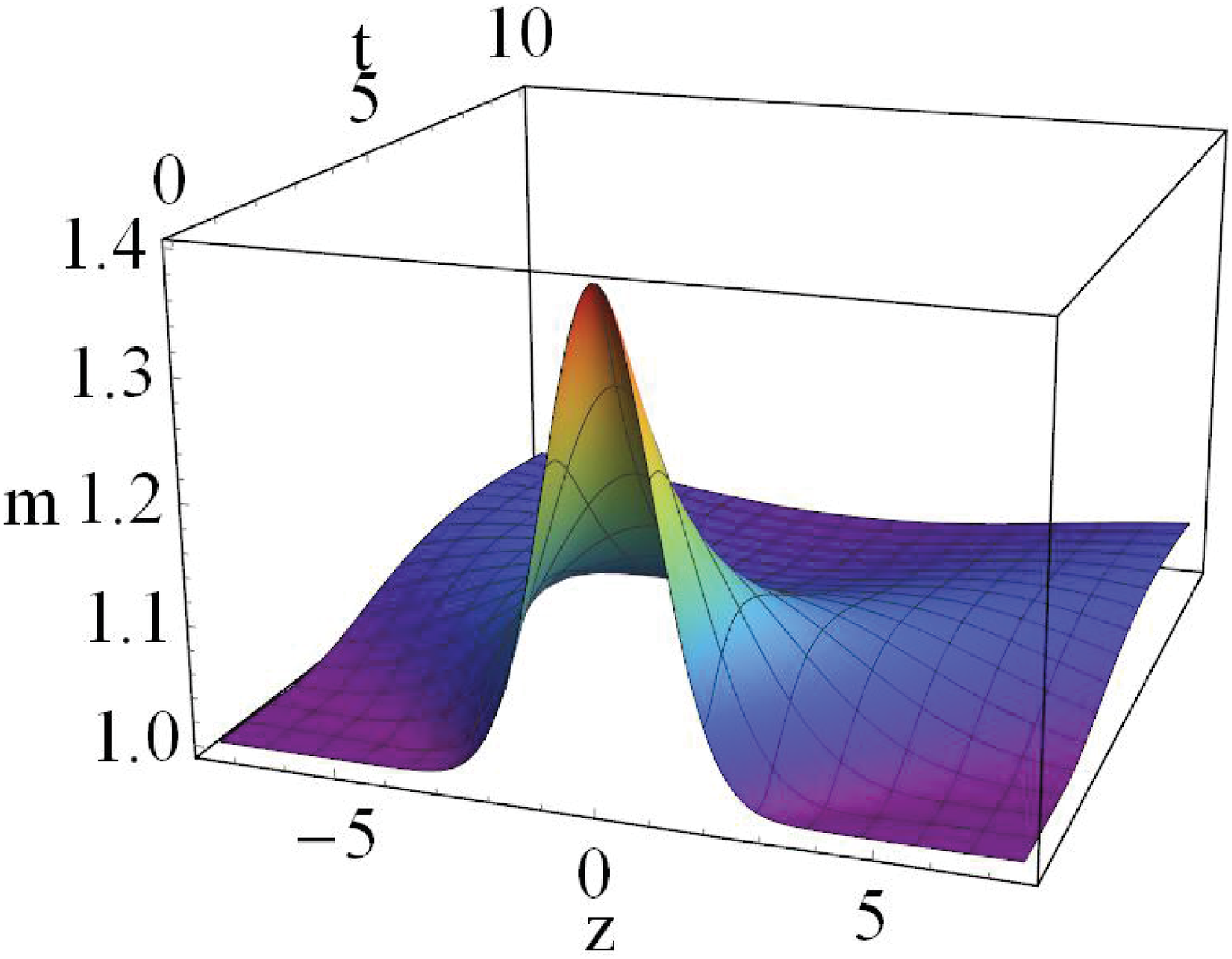} &
			\includegraphics[height=4.5cm]{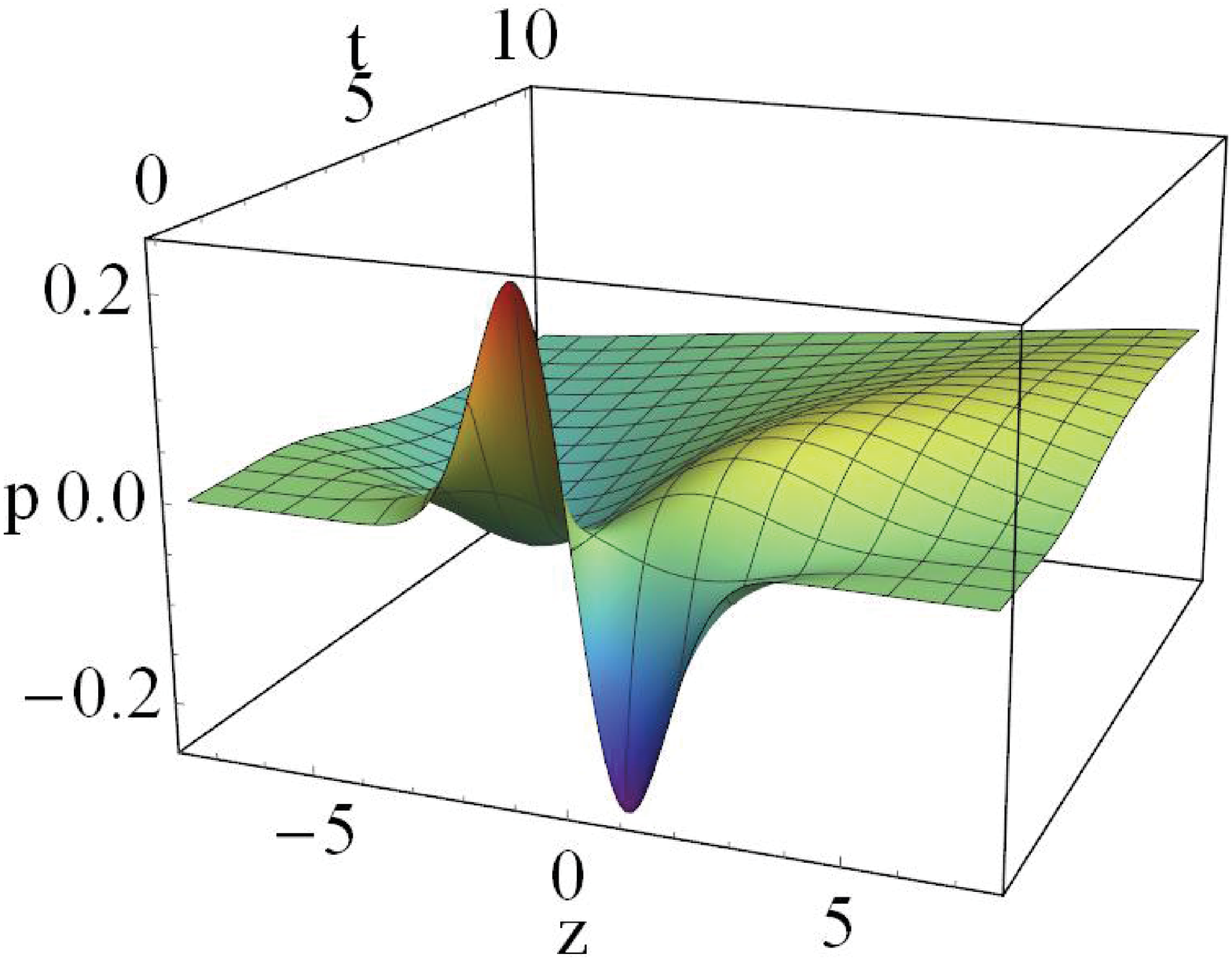} \\
			(c) & (d) \\
			\includegraphics[height=4cm]{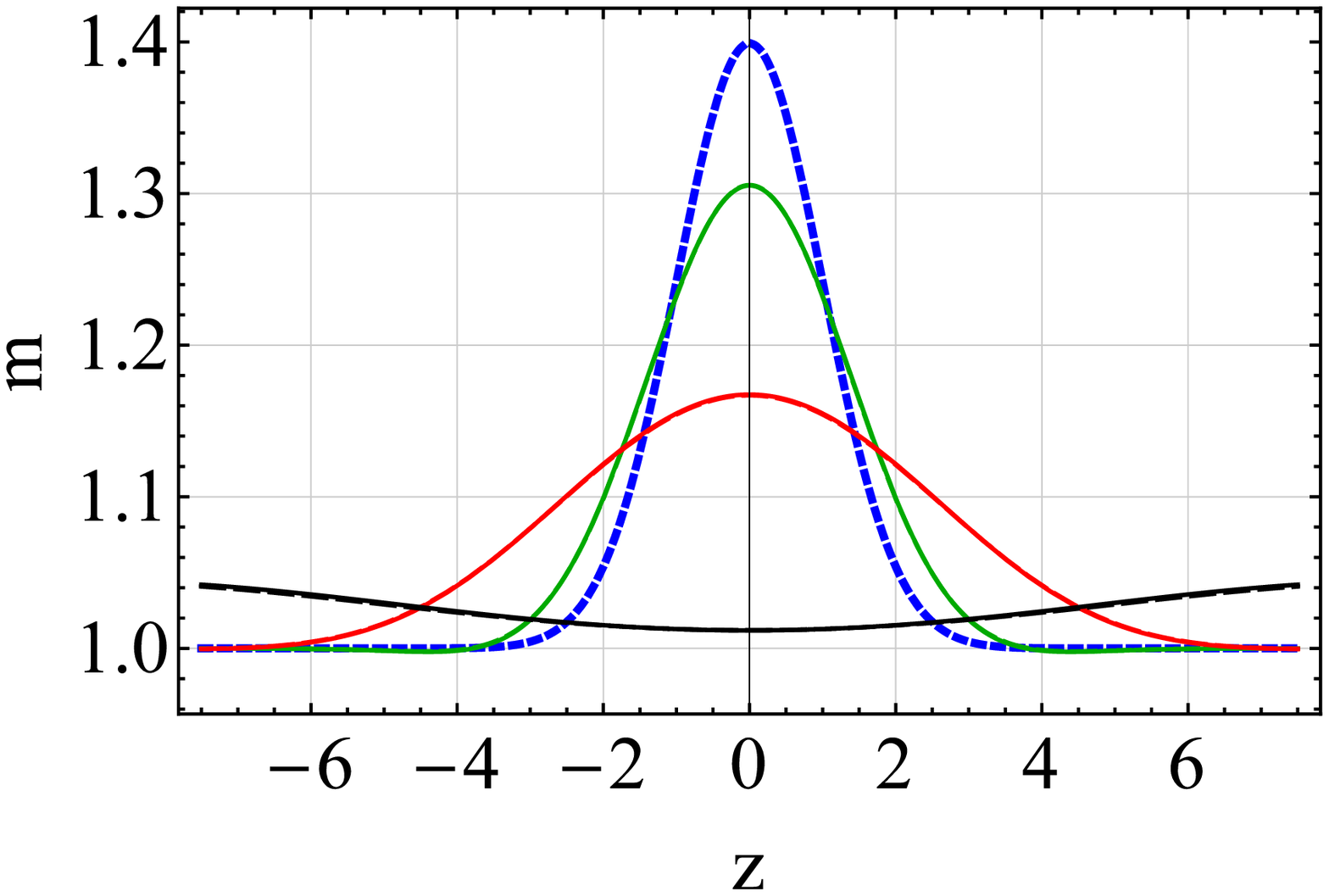} &
			\includegraphics[height=4cm]{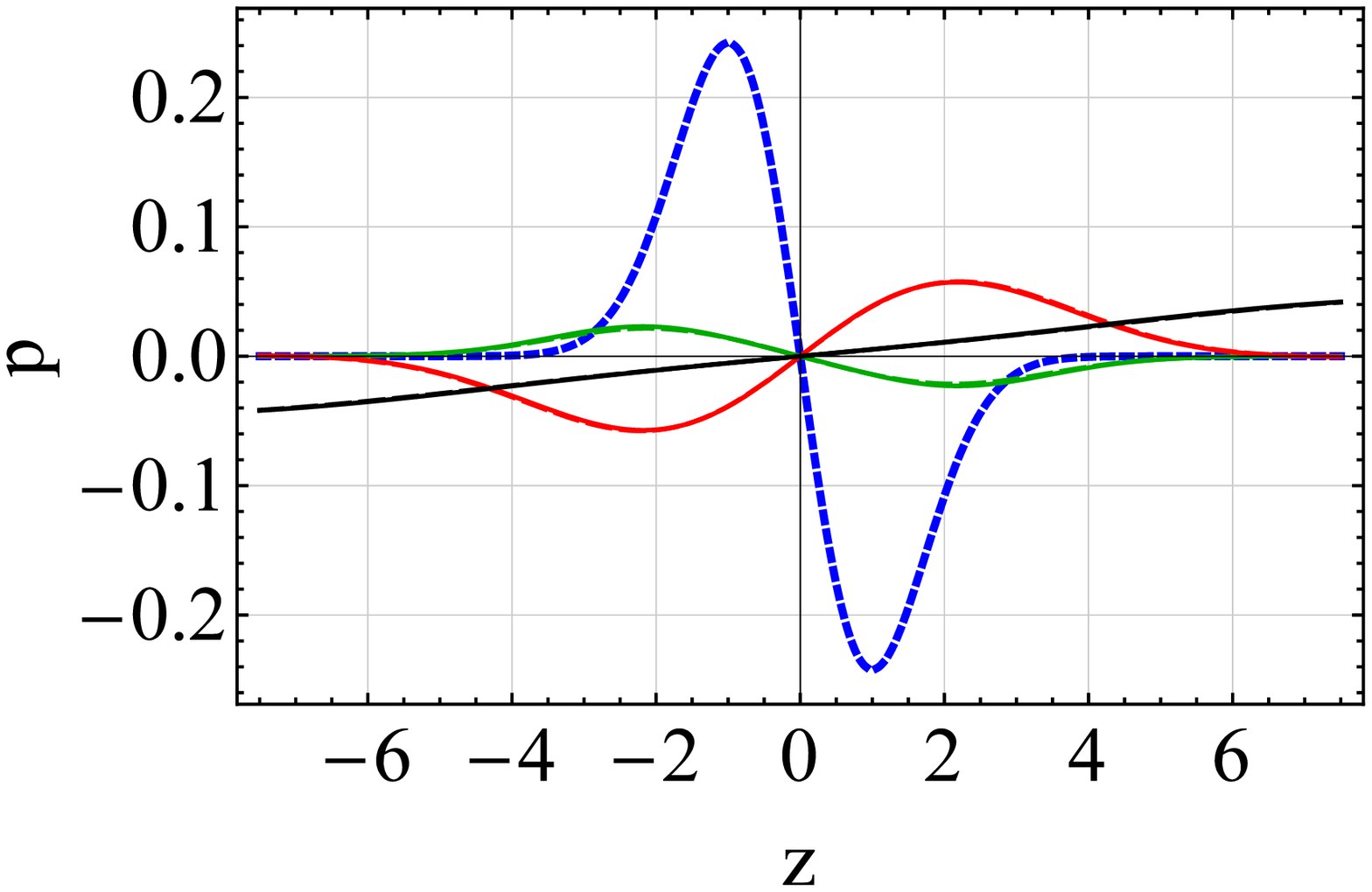} \\
		\end{tabular}
\caption{{\small {\sf Three-dimensional plots of (a) $1+m_1(t,z)$ and (b) $p_1(t,z)$, given by Eq.~\eqref{mp1_gauss_ads} with $\beta=b=1, \; z_0=0$. The comparison between first-order solution (c) $1+m_1(t,z)$ (resp.\ (d) $p_1(t,z)$) and full solution $m(t,z)$ (resp.\ $p(t,z)$) obtained by solving Eqs.~\eqref{eom1} and \eqref{eom2_ads} numerically. Snapshots of (c) $1+m_1(t,z)$ and (d) $p_1(t,z)$ at $t=0$ (blue-dashed), $t=0.67$ (green-solid), $t=2.0$ (red-solid), and $t=10$ (black-solid). Non-perturbative solutions obtained by solving Eqs.~\eqref{eom1} and \eqref{eom2_ads} numerically are drawn by (green, red, and black) dashed curves too, but can be hardly distinguished from the first-order solutions.}}}
\label{fig:gaussAdS}
\end{center}
\end{minipage}
	\end{center}
\end{figure}

\subsubsection{First-order solutions}

For the initial perturbation given by the Gaussian wave packet, Eqs.~\eqref{ic_gauss1} and \eqref{ic_gauss2}, one can compute the following quantities
\begin{align}
		{\cal F}^{-1}[ e^{{\mathsf s}_\sigma (k) t} \bar{m}_1 (0,k) ]
		&=
		\frac{\beta}{\sqrt{ 2\pi ( b^2+2t ) }}
		\exp
		[
			- \frac{ t^2 + (z-z_0)^2 }{ 2(b^2+2t) }
			-
			\sigma \frac{ t(z-z_0) }{ b^2+2t }
		],
\\
	{\cal F}^{-1}[ e^{{\mathsf s}_\sigma (k) t} \bar{p}_1 (0,k) ]
		&=
		-\frac{\beta ( z-z_0+\sigma t ) }{\sqrt{ 2\pi ( b^2+2t )^3 }}
		\exp
		[
			- \frac{ t^2 + (z-z_0)^2 }{ 2(b^2+2t) }
			-
			\sigma \frac{ t(z-z_0) }{ b^2+2t }
		].
\end{align}
Substituting these quantities into Eq.~\eqref{sol2_ads}, we obtain the first-order solution,
\be
		\left(
		\begin{array}{c}
			m_1 (t,z) \\
			p_1 (t,z) \\
		\end{array}
	\right)
	=
	\beta
	\sqrt{
		\frac{  (b^2+3t)^2 - (z-z_0)^2  }{ 2\pi (b^2+2t)^3 }
	} 	
	\exp[ - \frac{ t^2 + (z-z_0)^2 }{ 2(b^2+2t) } ]
	\left(
		\begin{array}{c}
			\displaystyle \cosh[ \frac{ t(z-z_0) }{ b^2+2t } - \Xi ] \\
			\displaystyle \sinh[ \frac{ t(z-z_0) }{ b^2+2t } - \Xi ] \\
		\end{array}
	\right),
\label{mp1_gauss_ads}
\ee
where
\be
	\cosh \Xi 
	:=
	\frac{ b^2+3t }{ \sqrt{ (b^2+3t)^2-(z-z_0)^2 } },
\;\;\;
	\sinh \Xi
	:=
	\frac{ z-z_0 }{ \sqrt{ (b^2+3t)^2-(z-z_0)^2 } }.
\ee

Like the diffusion phenomenon of a Gaussian wave packet according to an ordinary diffusion equation, solutions \eqref{mp1_gauss_ads} have temporal decay factor, which behaves as $  \frac{1}{\sqrt{ b^2+2t }} $, and spatial decay factor $\exp[ - \frac{ (z-z_0)^2 }{ 2(b^2+2t) } ]$. What crucially different from the ordinary diffusion is that the solution temporally damps rapidly due to factor $ \exp[ -\frac{ t^2 }{ 2(b^2+2t) } ] $.

Three-dimensional plots of $1+m_1(t,z)$ and $p_1(t,z)$ are presented in Figs.~\ref{fig:gaussAdS}(a) and \ref{fig:gaussAdS}(b), respectively. One can observe the fast damping described above. In addition, snapshots of $1+m_1(t,z)$ and $p_1(t,z)$ at selected moments, compared with numerical solutions, are presented in Figs.~\ref{fig:gaussAdS}(c) and \ref{fig:gaussAdS}(d), respectively. Compared with the full numerical solutions, which are obtained by directly solving original equations \eqref{eom1} and \eqref{eom2_ads} with the same initial conditions, {\it i.e.}, $m(0,z)=1+m_1(0,z)$ and $p(0,z)=p_1(0,z)$, one can observe that the first-order solution completely captures the full solution throughout the time domain considered. In other words, the higher-order perturbations are negligible, meaning that the $\epsilon$-expansion Eqs.~\eqref{expansion1} and \eqref{expansion2} converges rapidly for this example.

\subsection{Shock wave}
\label{sec:shock}

Here, let us consider the step-function like shock as the initial perturbation to the asymptotically AdS black brane. The propagation of this kind of shock is known as the Riemann problem in fluid mechanics. This classic problem attracts attentions recently in relativistic hydrodynamics since it makes us understand the non-equilibrium physics of quantum field theories. See the introduction of Ref.~\cite{Herzog:2016hob} for a brief but nice review for the recent development.

Assume that the black brane is given $O(\epsilon)$ perturbation as follows,
\begin{gather}
	m_1(0,z) = \alpha \;  {\rm sgn}(z),
\;\;\;
	p_1(0,z) = 0,
\label{shockic}
\end{gather}
where $\alpha$ is a real constant and sgn denotes the sign function,
\be
	{\rm sgn}(z) :=
	\begin{cases}
		-1 & (z<0) \\
		0	& (z=0) \\
		+1 & (z>0) \\
	\end{cases}.
\label{sgn}
\ee
The assumption that $p_1$ initially vanishes is adopted to reproduce a situation considered in Ref.~\cite{Herzog:2016hob} (see the left panel of Fig.~5 in \cite{Herzog:2016hob}). The Fourier transformation of the above initial conditions are
\be
	\bar{m}_1 (0,k) = - \frac{ 2i \alpha }{ k },
\;\;\;
	\bar{p}_1 (0,k) = 0.
\label{mpbar_shock_ads}
\ee

\begin{figure}[tb]
	\begin{center}
\begin{minipage}[c]{0.9\textwidth}
\linespread{0.85}
\begin{center}
		\setlength{\tabcolsep}{ 10 pt }
		\begin{tabular}{ cc }
					(a) & (b) \\
		\includegraphics[height=3.5cm]{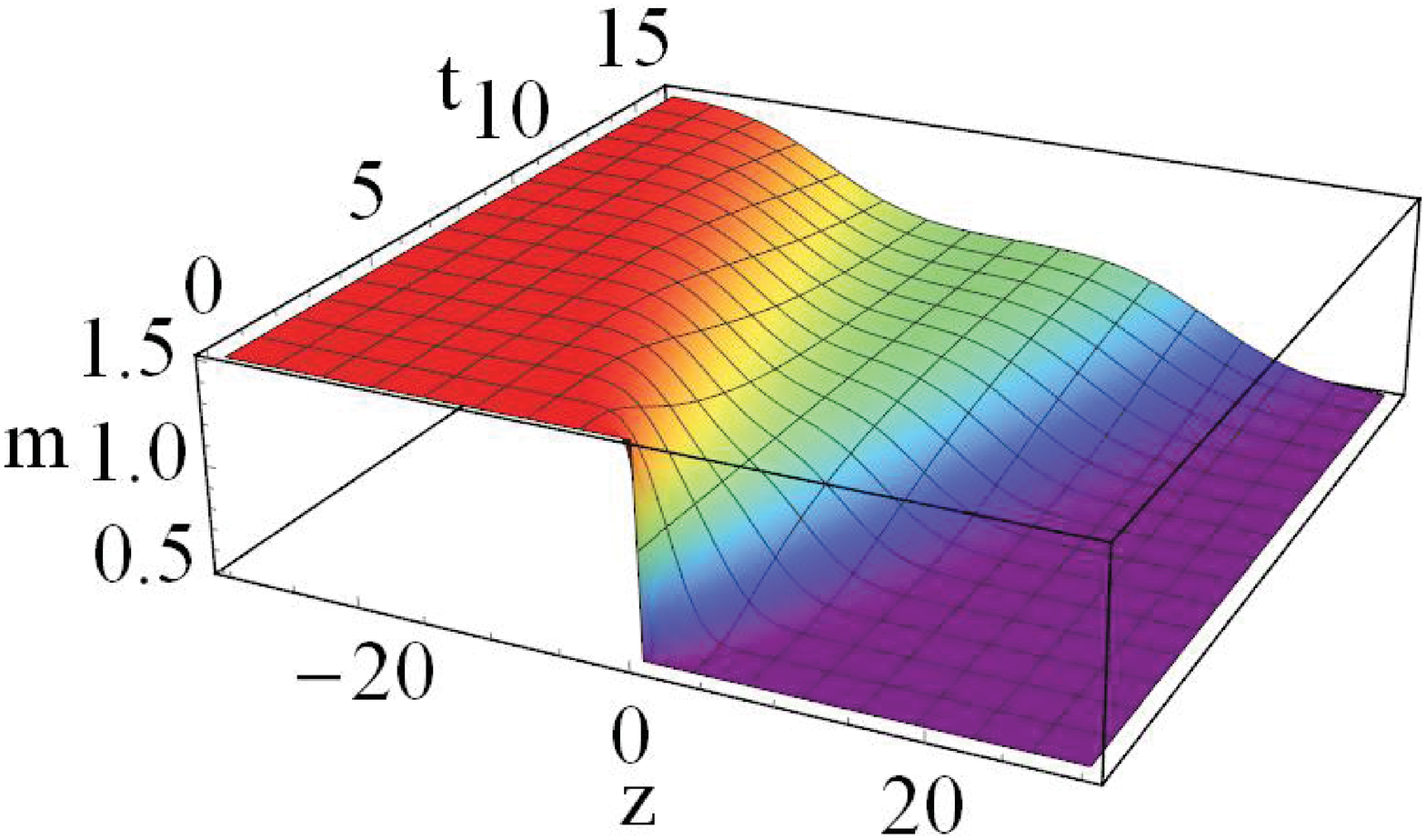}  &
		\includegraphics[height=3.5cm]{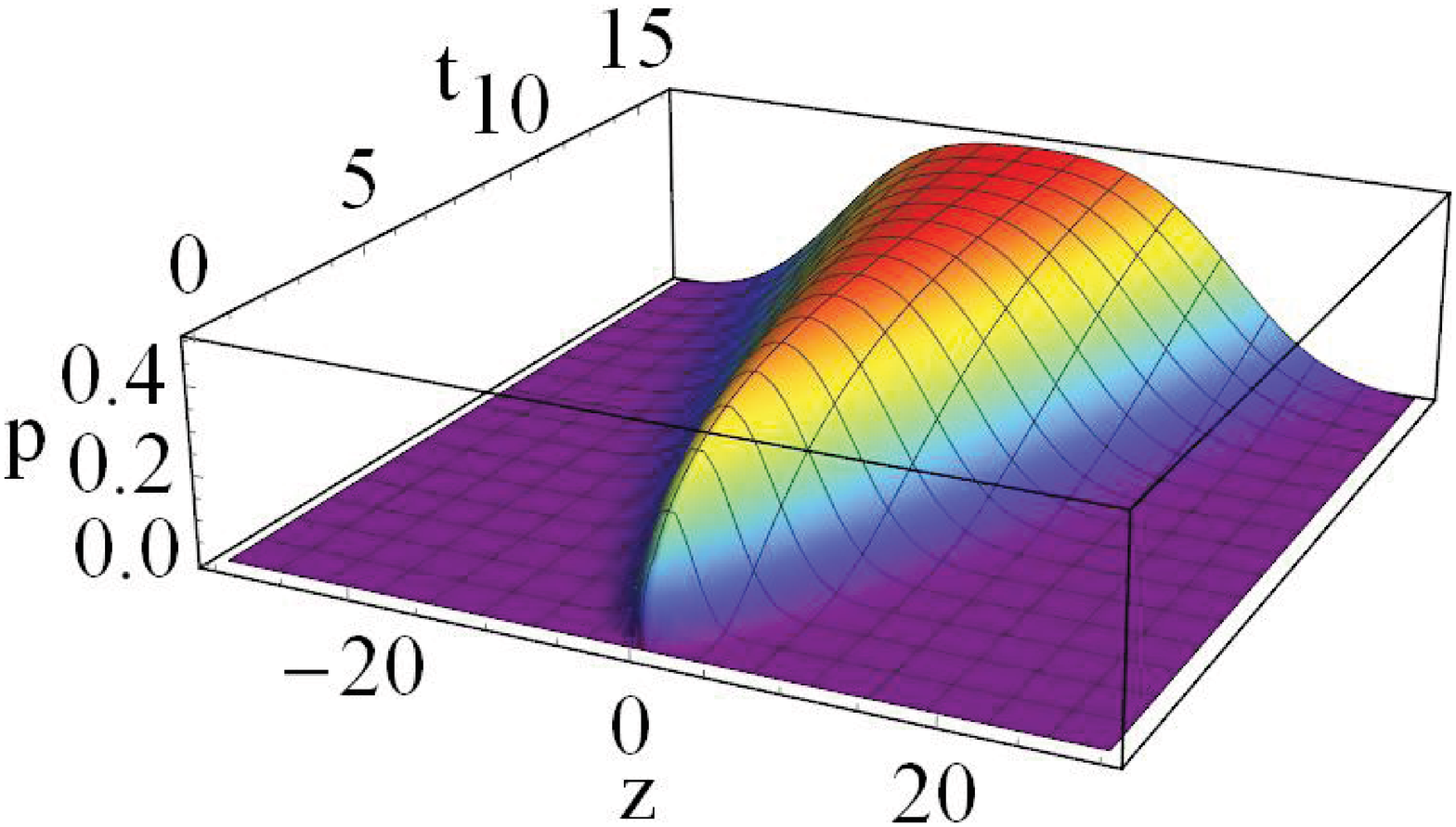} \\
					(c) & (d) \\
		\includegraphics[height=4cm]{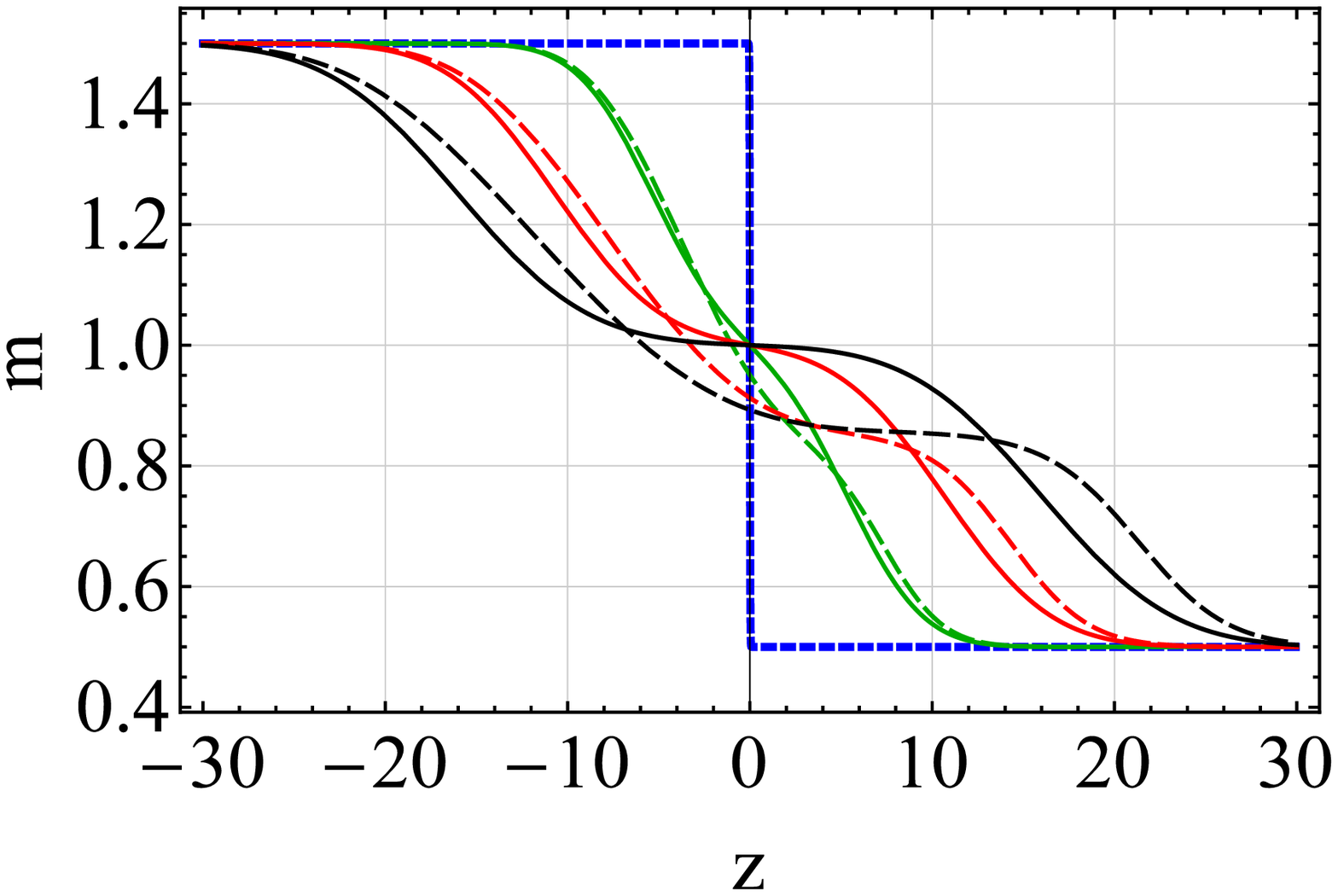}  &
		\includegraphics[height=4cm]{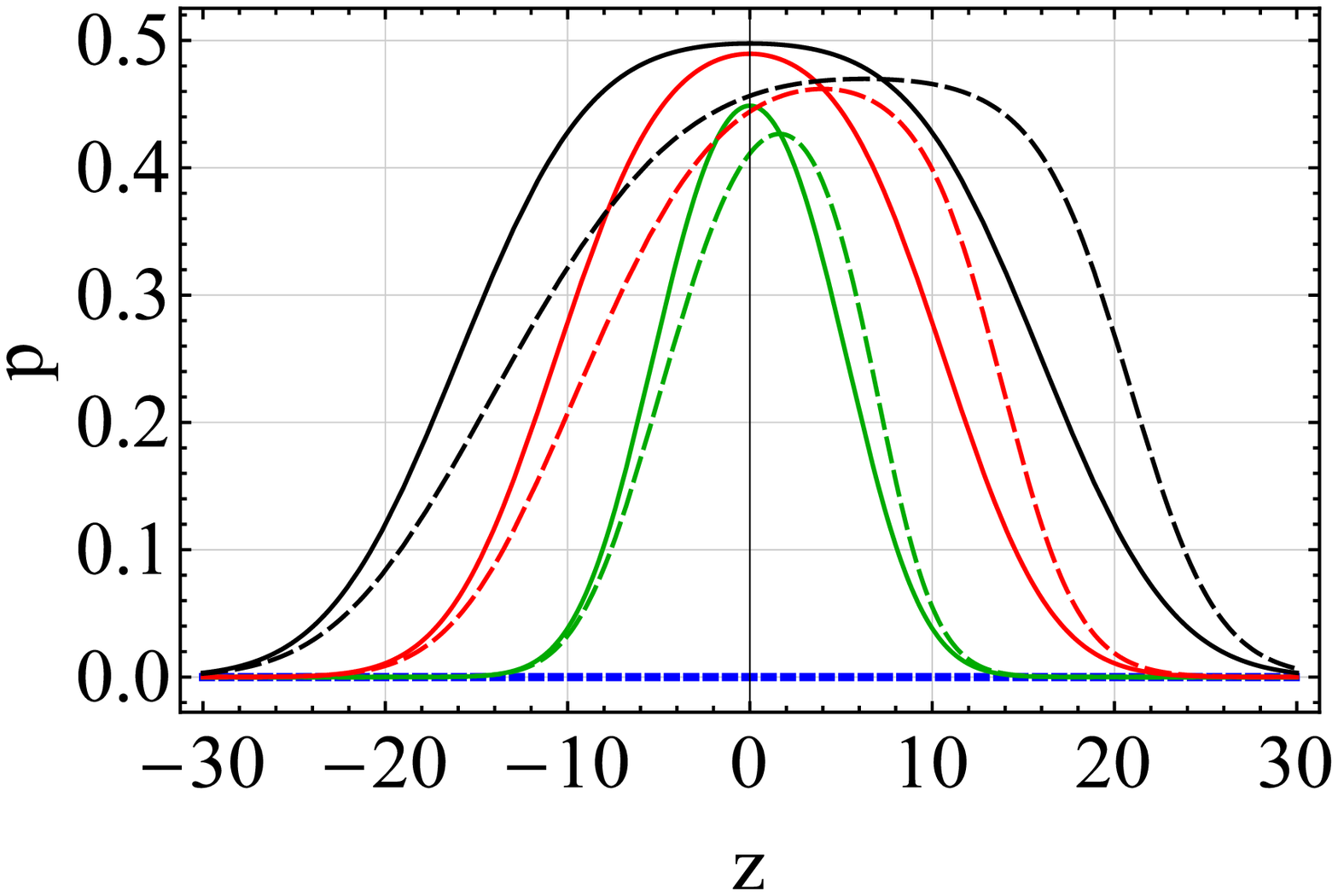} \\
		\end{tabular}
	\caption{{\small {\sf Three dimensional plots of (a) $1+m_1(t,z)$ and (b) $p_1(t,z)$, given by Eqs.~\eqref{shockm1} and \eqref{shockp1} with $\alpha=-1/2$, respectively. The comparison between first-order solution (c) $1+m_1(t,z)$ (resp.\ (d) $p_1(t,z)$) and full-order numerical solution $m(t,z)$ (resp.\ $p(t,z)$), obtained by solving Eqs.~\eqref{eom1} and \eqref{eom2_ads}. The blue-dashed curve represents initial configuration $m(0,z)$ (resp.\ $p(0,z)$). The green, red, and black solid curves represent the first-order solutions at $t=7.3$, $14$, and $22$, respectively. The (green, red, and black) dashed curve represents the full numerical solution at the corresponding time.}}}
\label{fig:shock}
\end{center}
\end{minipage}
	\end{center}
\end{figure}

\subsubsection{First-order solutions}

Using Eqs.~\eqref{dispersion_ads} and \eqref{mpbar_shock_ads}, one obtains
\be
	{\cal F}^{-1}[ e^{ {\mathsf s}_{\sigma} (k)t  } \bar{m}_1 (0,k)]
	=
	\alpha \; {\rm erf}( \frac{ \sigma t + z }{ 2\sqrt{t} } ),
\;\;\;
	{\cal F}^{-1}[ e^{ {\mathsf s}_{\sigma} (k)t  } \bar{p}_1 (0,k)] = 0,
\label{Fmp_shock}
\ee
where ${\rm erf}(x):=\frac{2}{\sqrt{\pi}} \int_0^x e^{-x^2}dx$ is the Gauss error function.  Here, we have used the following formula
\be
	\int_{-\infty}^\infty \frac{ e^{-a(k-ib)^2 } }{ k } dk
	=
	i \pi e^{ab^2} {\rm erf} (\sqrt{a} b^2),
\;\;\;
	a>0, \; b \in {\mathbb R}.
\ee
Substituting Eq.~\eqref{Fmp_shock} into Eq.~\eqref{sol2_ads}, one obtains the first-order solution,
\begin{align}
	m_1(t,z)
	&=\frac{\alpha}{2}
	\left[
		{\rm erf}(\frac{ t+z }{ 2\sqrt{t} }) - {\rm erf}(\frac{ t-z }{ 2\sqrt{t} } )
	\right],
\label{shockm1}
\\
	p_1(t,z)
	&= - \frac{\alpha}{2}
	\left[
		{\rm erf}(\frac{ t-z }{ 2\sqrt{t} }) + {\rm erf}(\frac{ t+z }{ 2\sqrt{t} })
	\right].
\label{shockp1}
\end{align}
Using the fact that the Gauss error function is an odd function, one can immediately show that $m_1(t,-z)=-m_1(t,z)$ and $p_1(t,-z)=p_1(t,z)$ hold. Namely, $m_1(t, z)$ and $p_1(t,z)$ are spatially odd and even functions, respectively.

In Fig.~\ref{fig:shock}, three-dimensional plots of $1+m_1(t,z)$ and $p_1(t,z)$, where the parameter is chosen as $\alpha=-1/2$, and their snapshots at selected moments are presented, compared with the non-perturbative numerical solutions obtained by directly solving Eqs.~\eqref{eom1} and \eqref{eom2_ads} with the initial condition similar to Eq.~\eqref{shockic}. Since the sign function is difficult to treat in numerical computations, we replace ${\rm sgn}(z)$ by $ \tanh ( c z )$ with a large positive number $c$ (say, $c=50$), based on the fact that $\lim_{c \to +\infty} \tanh(cz)={\rm sgn}(z)$.

In Fig.~\ref{fig:shock}, one can see that as soon as the dynamics begins, the discontinuity separates into two shock fronts moving to left and right, where the former and latter are called the rarefaction wave and shock wave, respectively~\cite{Herzog:2016hob}. The point is that the rarefaction and shock is interpolated by an expanding plateau region with a non-zero constant flux, called the non-equilibrium steady state (NESS).

While it is interesting that we obtained the semi-analytic results \eqref{shockm1} and \eqref{shockp1} describing the shock propagation and NESS, they are not satisfactory from some viewpoints. In Fig.~\ref{fig:shock}, one can observe that the full solution becomes asymmetric with respect to $z \to -z$, though the linear solutions \eqref{shockm1} and \eqref{shockp1} continue to be symmetric. Thus, for example, the value of $m$ at the NESS, which is always unity at $O(\epsilon)$, deviates between the first order and full solutions. These suggest that the higher-order perturbations are necessary to fill the above gaps. It seems impossible, however, to obtain the second-order solutions analytically since the first-order solution involves the error function. Thus, we will return to the Riemann problem in the next section as the example of the superposed sinusoidal waves.

\subsection{Superposed sinusoidal waves}
\label{sec:sin_ads}

We consider the initial condition which is the superposition of sinusoidal waves like Eqs.~\eqref{ic1}--\eqref{ic3}. For later purpose, however, let us assume the $O(\epsilon)$ initial momentum vanishes
\be
	p_1 (0,z) = 0,
\label{ic4}
\ee
which is assumed instead of Eq.~\eqref{ic2}. The first- and second-order solutions satisfying initial conditions~\eqref{ic1}--\eqref{ic3} are presented in Appendix~\ref{superpose_ads2}.

\subsubsection{First-order solutions}

For the initial perturbation which is the superposition of sinusoidal waves \eqref{ic1} and \eqref{ic4}, we obtain
\begin{gather}
	{\cal F}^{-1}[ e^{  {\mathsf s}_\sigma (k) t } \bar{m}_1 (0,k) ]
	=
	\frac12 \sum_{n=1}^N a_n
	[
		e^{ {\mathsf s}_\sigma (k_n) t } e^{ i( k_n z + \varphi_n ) }
		+
		e^{ {\mathsf s}_{- \sigma} (k_n) t } e^{ - i( k_n z + \varphi_n ) }
	],
\\
	{\cal F}^{-1}[ e^{  {\mathsf s}_\sigma (k) t } \bar{p}_1 (0,k) ]
	= 0,
\end{gather}
where we have used $ {\mathsf s}_\sigma (-k) = {\mathsf s}_{-\sigma} (k)$.
Substituting these results into Eq.~\eqref{sol2_ads} and using the concrete form of the dispersion relation ${\mathsf s}_{\sigma} (k)$, we obtain the  first-order solutions,
\begin{align}
	m_1(t,z)
	=&
	\sum_{n=1}^N 
	a_n e^{-k_n^2 t}
	\cos ( k_n t  )
	\cos ( k_n z + \varphi_n ),
\label{m1_ads}
\\
	p_1(t,z)
	=&
	\sum_{n=1}^N 
	a_n e^{-k_n^2 t}
	\sin ( k_n t )
	\sin ( k_n z + \varphi_n ).
\label{p1_ads}
\end{align}

These results \eqref{m1_ads} and \eqref{p1_ads} tell us that the initial perturbation necessarily exhibits the damped oscillation for arbitrary non-zero wave number $k_n \in {\mathbb R} \setminus \{ 0 \}$ ($n =1,2,\cdots , N$), showing the black brane to be linearly stable.

\subsubsection{Second-order solutions}

Since we assume that the initial values vanish at $O(\epsilon^2)$, $ m_2(0,z)=p_2(0,z)=0 $, we see from Eq.~\eqref{sol2_ads} that what to compute is the inverse Fourier transformation of the convolution between $ e^{ {\mathsf s}_\sigma (k) t } $ and the Fourier spectrum of source term $\bar{\psi}_2(t,k)$. Using Eqs.~\eqref{source2} and \eqref{p1_ads}, such a quantity is computed and written down in a simple form as
\begin{align}
	{\cal F}^{-1}[ e^{ {\mathsf s}_\sigma (k) t } \ast & \bar{ \psi }_2 (t,k) ]
	=
	- \frac{i}{8}
	\sum_{n=1}^N \sum_{n'=1}^N a_n a_{n'} k_{n'}
\nn
\\
&
\times
	\Big(
		F_{nn'}^{(\sigma)(+)} e^{i[ (k_n+k_{n'})z+(\varphi_n + \varphi_{n'}) ] } 
		+
		F_{nn'}^{(\sigma)(-)} e^{i[ (k_n-k_{n'})z+(\varphi_n - \varphi_{n'}) ]} 
\nn
\\
&
		-
		F_{nn'}^{(-\sigma)(-)} e^{-i[ (k_n-k_{n'})z+(\varphi_n - \varphi_{n'}) ]} 
		-
		F_{nn'}^{(-\sigma)(+)} e^{-i[ (k_n+k_{n'})z+(\varphi_n + \varphi_{n'}) ]} 
	\Big)
\label{convo_ads}
\end{align}
by defining a function of time,
\begin{align}
	F_{nn' }^{(\sigma)(\sigma') }
	:=&
	\frac{ 1 }{ {\mathsf s}_+(k_n) + {\mathsf s}_+(k_{n'}) - {\mathsf s}_\sigma ( k_n + \sigma' k_{n'} )}
	( e^{ [ {\mathsf s}_+(k_n) + {\mathsf s}_+(k_{n'}) ] t } - e^{ {\mathsf s}_\sigma (k_n +\sigma' k_{n'}) t } )
\nn
\\
	-&
	\frac{ 1 }{ {\mathsf s}_+(k_n) + {\mathsf s}_-(k_{n'}) - {\mathsf s}_\sigma ( k_n +\sigma' k_{n'} )}
	( e^{ [ {\mathsf s}_+(k_n) + {\mathsf s}_-(k_{n'}) ] t } - e^{ {\mathsf s}_\sigma (k_n +\sigma' k_{n'}) t } )
\nn
\\
	-&
	\frac{ 1 }{ {\mathsf s}_-(k_n) + {\mathsf s}_+(k_{n'}) - {\mathsf s}_\sigma( k_n +\sigma' k_{n'} )}
	( e^{ [ {\mathsf s}_-(k_n) + {\mathsf s}_+(k_{n'}) ] t } - e^{ {\mathsf s}_\sigma (k_n +\sigma' k_{n'}) t } )
\nn
\\
	+&
	\frac{ 1 }{ {\mathsf s}_-(k_n) + {\mathsf s}_-(k_{n'}) - {\mathsf s}_\sigma ( k_n +\sigma' k_{n'} )}
	( e^{ [ {\mathsf s}_-(k_n) + {\mathsf s}_-(k_{n'}) ] t } - e^{ {\mathsf s}_\sigma (k_n +\sigma' k_{n'}) t } ).
\label{f}
\end{align}
Substituting the above result \eqref{convo_ads} into Eq.~\eqref{sol2_ads}, we have the second-order solutions as
\begin{align}
m_2(t,z)
	=
	\frac{i}{8} \sum_{ n =1 }^n \sum_{n'=1}^N a_n a_{n'} k_{n'}
	\Big(
		[ & F_{nn'}^{(+)(+)} - F_{nn'}^{(-)(+)} ] 
		\cos [ (k_n+k_{n'})z + (\varphi_n + \varphi_{n'}) ]
\nn
\\
	+&
	 [ F_{nn'}^{(+)(-)} - F_{nn'}^{(-)(-)} ]  \cos [ (k_n-k_{n'})z + (\varphi_n - \varphi_{n'}) ]
	\Big),
\label{m2_pre_ads}
\\
p_2(t,z)
	=
	\frac18 \sum_{ n =1 }^n \sum_{n'=1}^N a_n a_{n'} k_{n'}
	\Big(
		[ & F_{nn'}^{(+)(+)} + F_{nn'}^{(-)(+)} ] 
		\sin [ (k_n+k_{n'})z + (\varphi_n + \varphi_{n'}) ]
\nn
\\
	+&
	 [ F_{nn'}^{(+)(-)} + F_{nn'}^{(-)(-)} ]  \sin [ (k_n-k_{n'})z + (\varphi_n - \varphi_{n'}) ]
	\Big) .
\label{p2_pre_ads}
\end{align}

Note that $m(t,z) = 1+m_1(t,z)+m_2(t,z)$ and $p(t,z) = p_1(t,z)+p_2(t,z)$ with Eqs.~\eqref{m1_ads}, \eqref{p1_ads}, \eqref{m2_pre_ads}, and \eqref{p2_pre_ads}, represent $O(\epsilon^2)$ approximate time evolution of the initial perturbation, which takes the form of superposed sinusoidal waves~\eqref{ic1}, \eqref{ic4}, and \eqref{ic3}. These approximate solutions are rather general in the sense that initial condition \eqref{ic1} is general.

Since the initial perturbations are assumed to vanish at $O(\epsilon^2)$, $m_2(0,z)=p_2(0,z)=0$, the above solutions contain only the contribution from the source term $\psi_2=-2p_1p_1'$. If one prepares for non-vanishing initial conditions at the second order, its contribution is simply added to the above solution, but such a contribution will exhibit no interesting behavior since it has the time dependence similar to that in $O(\epsilon)$ solution.

\subsubsection{Shock wave}

\begin{figure}[tb]
	\begin{center}
\begin{minipage}[c]{0.9\textwidth}
\linespread{0.85}
\begin{center}
			\setlength{\tabcolsep}{ 10 pt }
		\begin{tabular}{ cc }
				(a) & (b) \\
			\includegraphics[height=4cm]{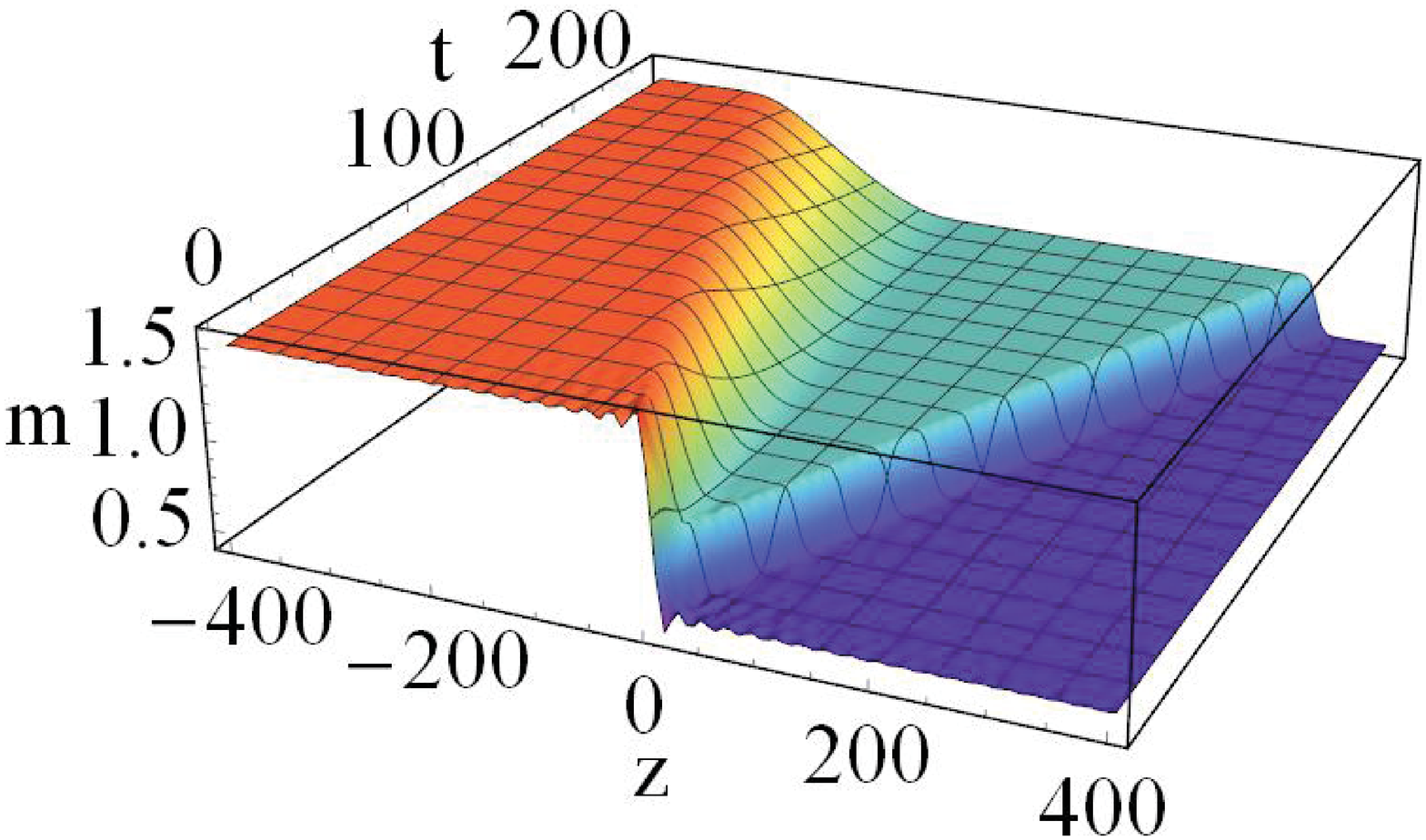} &
			\includegraphics[height=4cm]{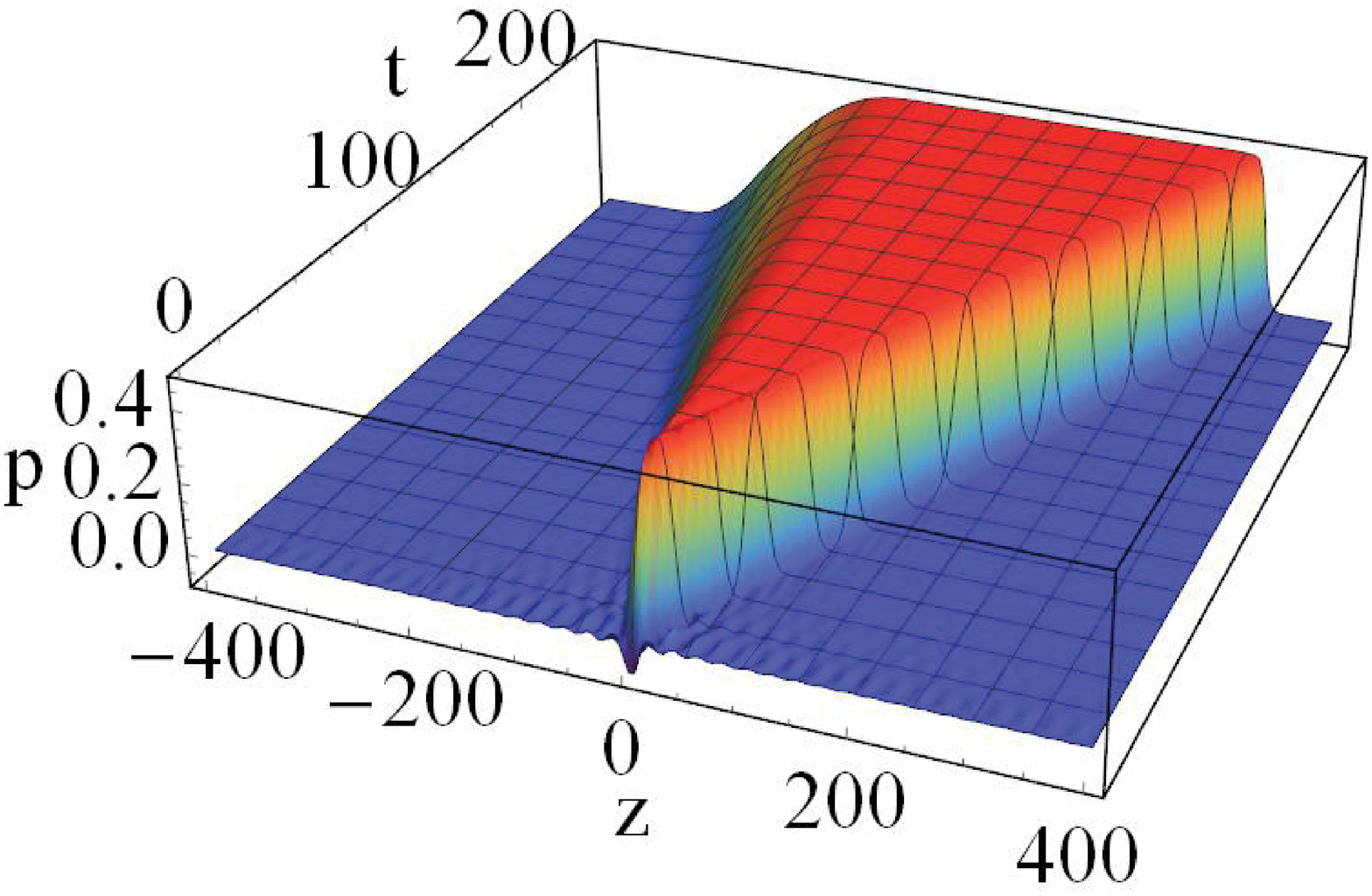} \\
				(c) & (d) \\
			\includegraphics[height=4cm]{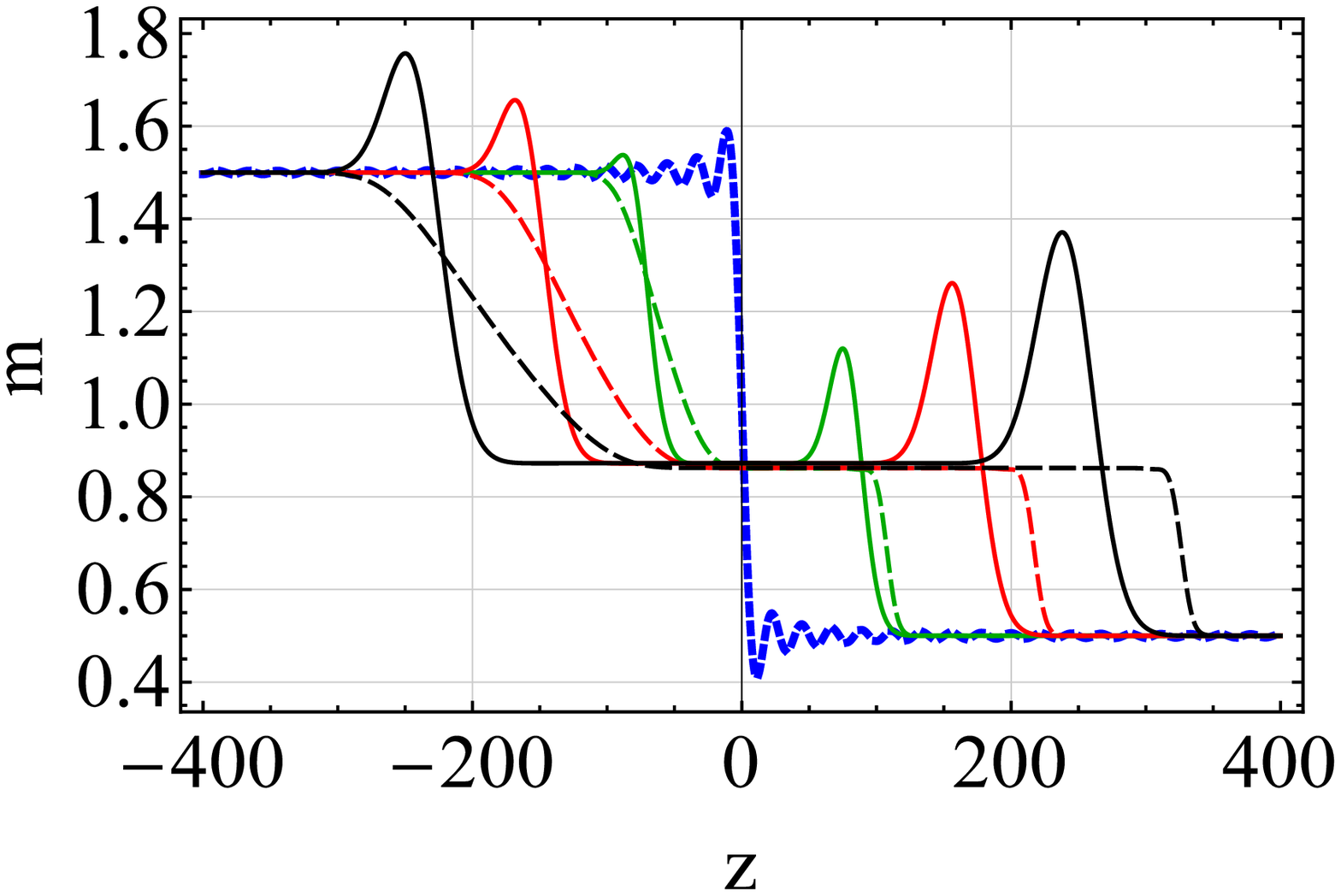} &
			\includegraphics[height=4cm]{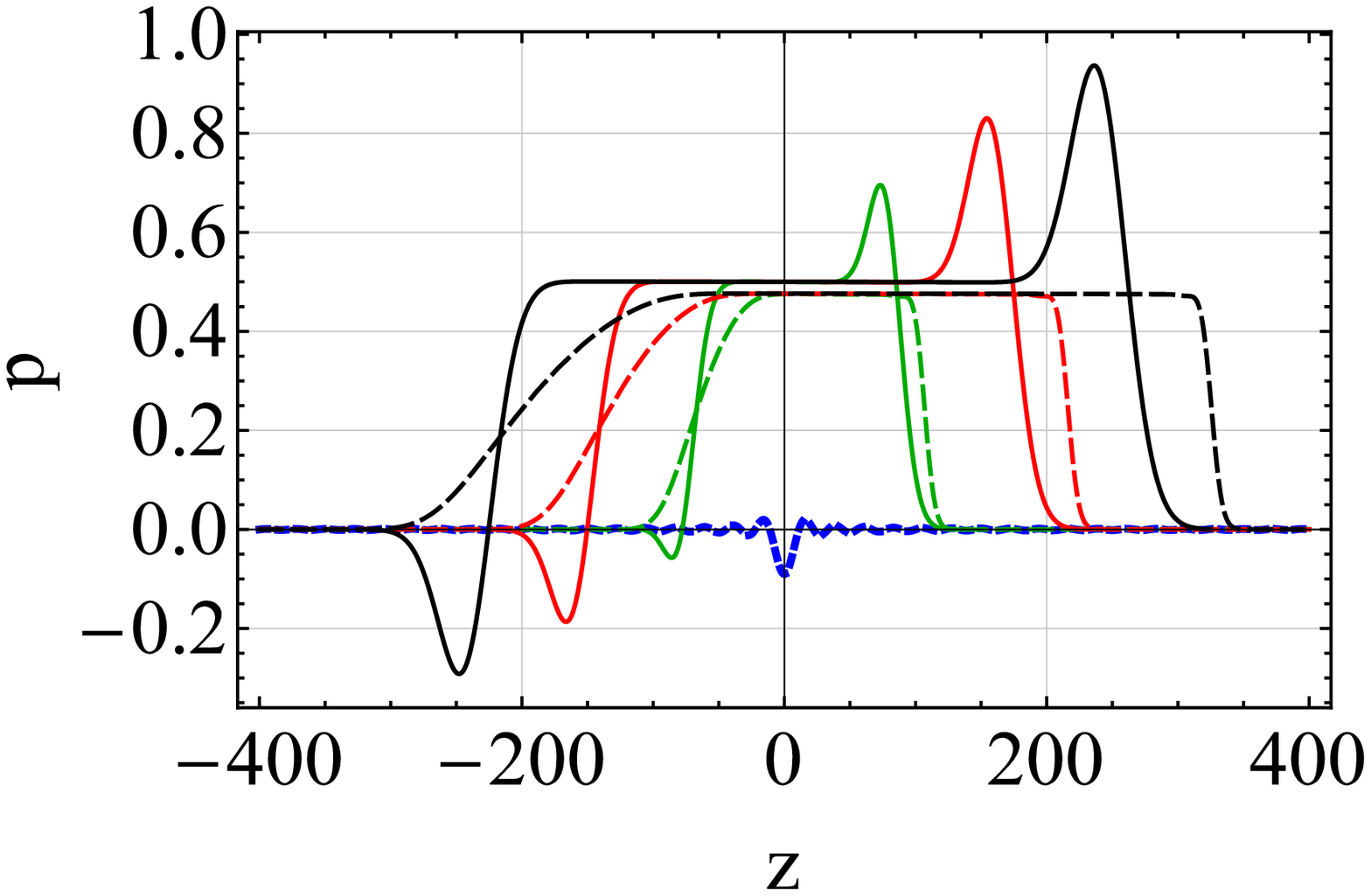} \\
		\end{tabular}
	\caption{{\small {\sf Three-dimensional plots of full-order numerical solution (a) $m(t,z)$ and (b) $p(t,z)$, obtained by solving Eqs.~\eqref{eom1} and \eqref{eom2_ads} with initial condition Eq.~\eqref{sgn} where ${\rm sgn}(z)$ is replaced by finite Fourier series of ${\rm sgn}_L(z)$ given by Eqs.~\eqref{sgnL1}--\eqref{sgnL3}. The comparison between second-order  solution (c) $1+m_1(t,z)+m_2(t,z)$ (resp.\ (d) $p_1(t,z)+p_2(t,z)$) and full-order numerical solution $m(t,z)$ (resp.\ $p(t,z)$), obtained by solving Eqs.~\eqref{eom1} and \eqref{eom2_ads}. The blue-dashed curve represents initial configuration $m(0,z)$ (resp.\ $p(0,z)$). The green, red, and black solid curves represent the second-order solutions at $t=83.3$, $167$, and $250$, respectively. The (green, red, and black) dashed curve represents the full numerical solution at the corresponding time.}}}
\label{fig:riemann}
\end{center}
\end{minipage}
	\end{center}
\end{figure}


We re-investigate the Riemann problem in Sec.~\ref{sec:shock}, by choosing appropriate parameters $(a_n, k_n, \varphi_n)$ in initial condition~\eqref{ic1}. In order to do so, the sign function, ${\rm sgn}(z)$, in initial condition~\eqref{shockic} is replaced by the following function,
\be
	{\rm sgn}_L(z) =
	\begin{cases}
		-1 & (-L < z<0) \\
		0	& (z = 0) \\
		+1 & (0 < z < L) \\
	\end{cases},
\ee
where $L$ is a positive constant and the periodic extension to entire ${\mathbb R}$ with period $2L$ is assumed. Since this function has the following Fourier series expansion,
\be
	{\rm sgn}_L(z)
	=
	\sum_{n=1}^\infty \frac{ 4 }{ (2n-1)\pi } \sin[ \frac{(2n-1)\pi }{L} z ],
\label{sgnL1}
\ee
the parameters in initial condition~\eqref{ic1} should be
\be
	a_n =    \frac{4}{(2n-1)\pi},
\;\;\;
	k_n = \frac{(2n-1)\pi}{L},
\;\;\;
	\theta_n = -\frac{\pi}{2}
\label{sgnL2}
\ee
for $n=1,2,\cdots, \; N$ with the limit $N \to \infty$. Taking large $L$ and focusing on the spatial region around the center $z=0$, there must be no difference in the dynamics during a finite interval of time between ones using ${\rm sgn}(z)$ and ${\rm sgn}_L(z)$, due to the (non-relativistic) causality encoded in the equations of motion.

In Figs.~\ref{fig:riemann}(a) and \ref{fig:riemann}(b), we present the three-dimensional plots of full-order numerical solutions of $m(t,z)$ and $p(t,z)$, respectively, starting from the initial condition $m(0,z)=1+m_1(0,z)$ and $p(0,z)=0$, where $m_1(0,z)$ is given by Eqs.~\eqref{ic1} and \eqref{sgnL2}. The rest parameters and the cutoff of $N$ are chosen as follow,
\be
	\alpha = - \frac12,
\;\;\;
	L=1000,
\;\;\;
	N=45.
\label{sgnL3}
\ee
In Figs.~\ref{fig:riemann}(c) and \ref{fig:riemann}(d), we compare these full solutions with the second-order solutions, {\it i.e.}, $1+m_1(t,z)+m_2(t,z)$ and $p_1(t,z)+p_2(t,z)$ provided by Eqs.~\eqref{m1_ads}, \eqref{p1_ads}, \eqref{m2_pre_ads}, and \eqref{p2_pre_ads} with initial conditions given by \eqref{ic1}, \eqref{sgnL2}, and \eqref{sgnL3}.

Unlike the $O(\epsilon)$ approximation in Sec.~\ref{sec:shock}, the $O(\epsilon^2)$ approximation presented here captures the spatially asymmetric features under the reflection $z \to -z$ of full solution. Furthermore, the error of values of $m$ and $p$ at the plateau of NESS between the $O(\epsilon^2)$ solution and full solutions are within a few percent ($1.2 \%$ and $5.0 \%$ for $m$ and $p$, respectively, for the above choice of parameters), which were inadequately large for the $O(\epsilon)$ approximation in Sec.~\ref{sec:shock}. This is by virtue of the second-order solutions, which appropriately take into account the `back-reaction' of the first-order solution through the source term. 

Due to the Gibbs phenomenon, there appears the oscillation near the jump in the initial shock. Such oscillations stem from the artificial cutoff of the Fourier series expansion. As the time proceeds, the `horns' as the remnant of Gibbs oscillations can be observed to propagate near the front of rarefaction and shock waves, and to be amplified. This artifact due to the cutoff will not be a problem when one treats the solution exactly.

\section{Conclusion}
\label{sec:conc}

In the large-$D$(dimension) limit of general relativity, the Einstein equations describing the horizon dynamics of asymptotically flat (resp.\ AdS) black branes are written in the form of coupled diffusion equations \eqref{eom1} and \eqref{eom2} (resp.\ \eqref{eom1} and \eqref{eom2_ads})~\cite{Emparan:2015gva,Herzog:2016hob}. While these equations are much simpler than the original Einstein equations, there is a non-linear term, making it difficult to solve exactly. Therefore, we have formulated the perturbation theory in this paper making it possible to obtain analytic results on the black-brane dynamics.

The metric functions $m(t,z)$ and $p(t,z)$, which represent the mass and momentum distributions in the direction of horizon $z$, respectively, were expanded around a uniform black-brane solution with formal small parameter $\epsilon$ as Eqs.~\eqref{expansion1} and \eqref{expansion2}. Then, the perturbative equations of motion were obtained, and solved order by order using the Laplace and Fourier transformation with respect to $t$ and $z$, respectively. As the result, the general form of solutions $m_\ell (t,z)$ and $p_\ell(t,z)$ in the flat (resp.~AdS) background were obtained~\eqref{sol2} (resp.~\eqref{sol2_ads}), in which the inverse Fourier transformation of initial spectra $\bar{m}_\ell (0,k)$ and $\bar{p}_\ell (0,k)$, and the spectrum of source term $\bar{\psi}_\ell(t,k)$, being a polynomial of lower-order perturbations, were left to be computed.

As the example of initial conditions of perturbation, the Gaussian wave packet was considered for both the asymptotically flat and AdS black branes in Secs.~\ref{sec:gauss} and \ref{sec:gauss_ads}, respectively, and the first-order solutions were written down explicitly. The resulting dynamics from this initial condition was not surprising itself. Namely, the wave pack grows and damps rapidly for the asymptotically flat and AdS black branes, respectively, which is expected from their known stability. A remarkable point revealed by this example is that the first-order solution captures the features of full-order solution rather accurately even for a finite amplitude. Thus, the convergence of expansion by $\epsilon$ ({\it i.e.},\ amplitude) is rather rapid and can be used for finite-amplitude perturbations for a certain class of problems.

Only for the asymptotically AdS black brane, the step-function like initial condition was considered in order to investigate the shock propagation in Sec.~\ref{sec:shock}, and the first-order solution was explicitly written down. While the solution captures the emergence and propagation of NESS (non-equilibrium steady state) qualitatively, the first-order solution is not enough to reproduce the properties of NESS such as the values of metric functions and asymmetry of the full solution. 

The discretely superposed sinusoidal waves, which can be the Fourier series expansion of an arbitrary piecewise continuous periodic function, were considered for both the asymptotically flat and AdS black branes in Secs.~\ref{sec:sin} and \ref{sec:sin_ads}, respectively, and the first- and second-order solutions were written down explicitly. For the black brane in the flat background, the non-trivial feature of GL (Gregory-Laflamme) instability was revealed. Namely, the mode-mode coupling at the second order can make the perturbation grows even if the initial perturbation damps at the first order.  For the black brane in the AdS background, the shock propagation considered in Sec.~\ref{sec:shock} was re-investigated. Thanks to the second-order contribution, the values and asymmetric features of the full solution were reproduced, illustrating the usefulness of the formalism in this paper.

There are many things to do by applying and generalizing the formulation presented in this paper. For instance, (i) it would be interesting to investigate further why the sign of convergence to non-uniform black string (NUBS) of asymptotically flat black string does not appear even in the second-order perturbation. (ii) One is able to investigate the various type of shock-wave propagation such as those discussed in Ref.~\cite{Herzog:2016hob} if one slightly changes the ansatz of expansion \eqref{expansion1} and \eqref{expansion2}, which is straightforward. (iii) Including $1/D$ corrections and charges of background black branes~\cite{Emparan:2015hwa,Emparan:2016sjk,Rozali:2016yhw} would increase the problems to be worked by our formalism.

\subsection*{Acknowledgments}
The author would like to thank R.\ Emparan, R.\ Suzuki, and K.\ Tanabe for useful discussion and comments during The Spanish-Portuguese Relativity Meetings held at Lisbon  (12--15th, Sep.~2016), and T.~Torii during his stay at Akita (12--17th, Feb.\ 2017). This work was supported by JSPS KAKENHI Grant Number 15K05086.

\appendix

\section{Sinusoidal waves in AdS with $p_1(0,z)=\pd_z m_1(0,z)$}
\label{superpose_ads2}

For the initial perturbation to the asymptotically AdS black branes which is the superposition of sinusoidal waves \eqref{ic1}--\eqref{ic3}, we obtain
\begin{align}
	m_1(t,z)
	=&
	\frac12 \sum_{n=1}^N a_n
	[
		(1 -i k_n) e^{ {\mathsf s}_+ (k_n) t }  + (1 +i k_n) e^{ {\mathsf s}_- (k_n) t }
	] \cos ( k_n z + \varphi_n ),
\label{m1_ads_p1=0}
\\
	p_1(t,z)
	=&
	- \frac{i}{2} \sum_{n=1}^N a_n
	[
		(1 -i k_n) e^{ {\mathsf s}_+ (k_n) t }  - (1 + i k_n) e^{ {\mathsf s}_- (k_n) t }
	] \sin ( k_n z + \varphi_n ).
\label{p1_ads_p1=0}
\end{align}
Using the concrete form of the dispersion relation ${\mathsf s}_{\sigma} (k)$, we obtain the following form of first-order solution,
\begin{align}
	m_1(t,z)
	=&
	\sum_{n=1}^N 
	\sqrt{ 1+k_n^2 } a_n e^{-k_n^2 t}
	\cos ( k_n t + \vartheta_n )
	\cos ( k_n z + \varphi_n ),
\label{m1_ads_p1=0_a}
\\
	p_1(t,z)
	=&
	\sum_{n=1}^N 
	\sqrt{ 1+k_n^2 } a_n e^{-k_n^2 t}
	\sin ( k_n t + \vartheta_n )
	\sin ( k_n z + \varphi_n ),
\label{m1_ads_p1=0_b}
\end{align}
where
\be
	\cos \vartheta_n := \frac{1}{\sqrt{1+k_n^2}},
\;\;\;
	\sin \vartheta_n := - \frac{ k_n }{\sqrt{1+k_n^2}}.
\ee
Equations \eqref{convo_ads}, \eqref{m2_pre_ads}, and \eqref{p2_pre_ads} hold just by replacing  $F_{nn' }^{(\sigma)(\sigma') } $ for the following $\tilde{F}_{nn' }^{(\sigma)(\sigma') }$,
\begin{align}
	\tilde{F}_{nn' }^{(\sigma)(\sigma') }
	:=&
	\frac{ ( 1 -i k_n )( 1-ik_{n'} ) }{ {\mathsf s}_+(k_n) + {\mathsf s}_+(k_{n'}) - {\mathsf s}_\sigma ( k_n + \sigma' k_{n'} )}
	( e^{ [ {\mathsf s}_+(k_n) + {\mathsf s}_+(k_{n'}) ] t } - e^{ {\mathsf s}_\sigma (k_n +\sigma' k_{n'}) t } )
\nn
\\
	-&
	\frac{ ( 1-ik_n )( 1+ik_{n'} ) }{ {\mathsf s}_+(k_n) + {\mathsf s}_-(k_{n'}) - {\mathsf s}_\sigma ( k_n +\sigma' k_{n'} )}
	( e^{ [ {\mathsf s}_+(k_n) + {\mathsf s}_-(k_{n'}) ] t } - e^{ {\mathsf s}_\sigma (k_n +\sigma' k_{n'}) t } )
\nn
\\
	-&
	\frac{ ( 1+ik_n )( 1-ik_{n'} ) }{ {\mathsf s}_-(k_n) + {\mathsf s}_+(k_{n'}) - {\mathsf s}_\sigma( k_n +\sigma' k_{n'} )}
	( e^{ [ {\mathsf s}_-(k_n) + {\mathsf s}_+(k_{n'}) ] t } - e^{ {\mathsf s}_\sigma (k_n +\sigma' k_{n'}) t } )
\nn
\\
	+&
	\frac{ ( 1+ik_n )( 1+ik_{n'} ) }{ {\mathsf s}_-(k_n) + {\mathsf s}_-(k_{n'}) - {\mathsf s}_\sigma ( k_n +\sigma' k_{n'} )}
	( e^{ [ {\mathsf s}_-(k_n) + {\mathsf s}_-(k_{n'}) ] t } - e^{ {\mathsf s}_\sigma (k_n +\sigma' k_{n'}) t } ).
\label{f2}
\end{align}



\end{document}